\pdfoutput=1   
\documentclass[usenatbib,usegraphicx,fleqn]{ctextemp_mn2e} 
\usepackage{amsmath,amssymb,bm,color,multirow,multicol,eufrak,relsize}
\usepackage{charter}   
\usepackage{color}

\newcommand{\fig}[1]{Fig. \ref{#1}}

\newcommand{\slfrac}[2]{\left.#1\middle/#2\right.}

\usepackage{ulem}

\def\prodi#1#2{\frac{d #1}{d #2}}

\usepackage{color}

\title[Effect of magnetic field on planetary migration]{The effects of a magnetic field on planetary migration in laminar and turbulent discs}

\author[M.L. Comins et al.]{\parbox{\textwidth}{
Megan L.~Comins$^{1}$\thanks{E-mail of corresponding author: \texttt{mcomins@astro.cornell.edu}},
Marina M.~Romanova$^{1}$,
Alexander V.~Koldoba$^{2,3}$,
Galina V.~Ustyugova$^{4}$,
Richard V.E.~Lovelace$^{1,5}$}\vspace{0.4cm}\\
\parbox{\textwidth}{ 
$^{1}$Department of Astronomy, Cornell University, Ithaca, NY 14853-6801\\
$^{2}$Institute of Computer Aided Design RAS, $2^{nd}$ Brestskaya st., 19/18, Moscow, 123056, Russia\\
$^{3}$Moscow Institute of Physics and Technology, Dolgoprudny, Moscow Region, 141700, Russia\\
$^{4}$Keldysh Institute for Applied Mathematics, Moscow, Russia\\
$^{5}$Department of Applied and Engineering Physics, Cornell University, Ithaca, NY 14853-680
}}

\begin{document}

\maketitle

\begin{abstract}

We investigate the migration of low-mass planets ($5 M_{\oplus}$ and $20 M_{\oplus}$) in accretion discs threaded with a magnetic field using 
2D MHD code in polar coordinates. We observed that, in the case of a strong azimuthal magnetic field where the plasma parameter is 
$\beta\sim 1-2$, density waves at the magnetic resonances exert a positive torque on the planet and may slow down or reverse its migration.  However, 
when the magnetic field is weaker (i.e., the plasma parameter $\beta$ is relatively large), then non-axisymmetric density waves excited by the planet 
lead to growth of the radial component of the field and, subsequently, to development of the magneto-rotational instability,  such that the disc becomes
turbulent. Migration in a turbulent disc is stochastic, and the migration direction may change as such. To understand migration in a turbulent disc, both
the interaction between a planet and individual turbulent cells, as well as the interaction between a planet and ordered density waves, have been
investigated.

\end{abstract}


%
%

\section{Introduction}
\label{sec:intro}

The migration of protoplanets embedded in accretion discs has been extensively studied in recent years (see, e.g., review by
\citealt{KleyNelson2012}). A low-mass protoplanet excites density waves in the disc at the Lindblad resonances and migrates toward the star if no 
other torques are exerted (e.g., \citealt{GoldreichTremaine1979}). However, a corotation torque originating near the planet can slow or even reverse
the migration (e.g., \citealt{TanakaEtAl2002,Masset2002,MassetEtAl2006}). Another possible cause of slowed or reversed migration is the excitation 
of magnetic resonances in the disc when a relatively strong ($\beta \sim 1$\footnote{$\beta = \slfrac{8\pi P}{B^2}$, where $P$ and $B$ are the 
pressure and magnetic field in the disc, respectively}) azimuthal magnetic field is present \citep{Terquem2003}. These magnetic resonances can halt or
reverse the migration of the planet if the magnetic field is sufficiently strong and it has a steep gradient toward the star. The action of this mechanism
has been shown in numerical simulations by \citet{FromangEtAl2005}.

On the other hand, accretion discs threaded with a weak magnetic field are prone to the magneto-rotational instability (MRI;
\citealt{BalbusHawley1991,BalbusHawley1998}). Numerical simulations of planet migration in a turbulent disc have shown that the semimajor axis of the 
planet changes stochastically due to the planet's interaction with the turbulent cells in the disc. The migration rate can increase or decrease, and the 
direction of the migration can reverse \citep{NelsonPapaloizou2004}.

An initial goal of this paper was to more deeply investigate the effect of magnetic resonances on the migration of a low-mass embedded planet following 
the methods used by \citet{FromangEtAl2005}, albeit that we explored a larger parameter space with regards to the mass of the planet and the density 
distribution in the disc, as well as the magnetic field strength and its distribution in the disc.
While exploring this expanded parameter space, we noticed the development of MRI-driven turbulence in the disc in many cases. In particular, when the 
magnetic field is weak, non-axisymmetric density waves excited by the planet lead to perturbations in the initially-azimuthal magnetic field and, 
subsequently, to growth of the radial component of the magnetic field near the planet. The azimuthal field grows with time due to 
the stretching of the radial field lines by the differential rotation of the Keplerian disc. As a result, an MRI-type instability develops and propagates to 
larger distances from the planet, and the disc becomes turbulent. The planet interacts with the turbulent cells, and the torque associated with this 
interaction is much larger than the torque seen prior to the development of turbulence (i.e., the torque due to magnetic resonances). Therefore, we 
also studied the parameter space describing the transition from a laminar to a turbulent disc, as well as the migration of a planet in a turbulent disc in 
detail.

We first describe how the migration rate and direction of a low-mass planet are affected by the planet mass, as well as the distribution of surface 
density and magnetic field in the laminar disc. We also show how these parameters alter the time at which the onset of turbulence occurs in magnetized 
discs. Finally, to better understand the interaction between the planet and turbulent cells in the disc, we study the torque on the planet by (1) an 
ordered, low-amplitude density wave generated at the inner boundary and (2) a high-amplitude wave that is excited in a turbulent MHD disc.

The plan for this paper is as follows.  In \S \ref{sec:theoretical background} we overview the theory of the different sources of torque on the planet. In 
\S \ref{sec:model} we describe our physical model and numerical setup. In \S \ref{sec: parameter space} we describe our main parameters. We 
describe our test simulations in the hydrodynamical disc in \S \ref{sec:migration hydro} and simulations in a laminar magnetic disc in \S \ref{sec:magres 
migration}. Migration in turbulent discs is shown in \S \ref{sec: migration turbulent}, and migration under the influence of density waves is shown in \S 
\ref{sec:migration in waves}. We conclude in  \S \ref{sec:conclusions}.

%
%

\section{Theoretical background}
\label{sec:theoretical background}

In this section, we discuss the causes of  inward and outward migration in hydrodynamic and MHD discs.

\subsection{Migration in hydrodynamic discs}
\label{section:hydro_torque}

In the absence of a magnetic field in the disc, the migration of the planet is determined by the Lindblad and corotation torques 
\citep{GoldreichTremaine1979,Ward1986,Ward1997}. The Lindblad torque is generated when the planet excites $m$-armed waves in the disc with 
``orbital'' frequencies $\omega = m\Omega_{\rm p}$. The dispersion relation for these waves in a low-temperature, low-mass disc is 
$\left(\omega - m\Omega\right)^{2} = \kappa^{2},$ where $\kappa = \Omega_{\rm Kep}$ is the epicyclic frequency in a Keplerian disc. Substituting 
$\Omega(r)~=~\sqrt{{GM_{\star}}/{{r^{3}}}}$, the locations of the Lindblad resonances are
\begin{equation}
    r_{\rm LR} = r_{\rm p} \left(\frac{m \pm 1}{m} \right)^{2/3}.
\end{equation}
For each value of $m$ (except $m=1$), there is one resonance located closer to (farther from) the star than the planet called the inner (outer) Lindblad 
resonance. The inner Lindblad resonances exert a positive torque on the planet and ``push'' the planet outward, while the outer Lindblad resonances 
have the opposite effect. The total Lindblad torque (also known as the ``differential Lindblad torque'') determines the migration direction in the absence 
of other torques. The differential Lindblad torque is negative, resulting in overall inward migration of the planet if there are no other sources of torque.

However, there is typically a corotation torque exerted on the planet as well. The physics of the corotation torque and its effect on planet migration has 
been studied by number of authors (e.g., \citealt{PaardekooperMellema2006,BaruteauMasset2008,PaardekooperPapaloizou2008,
KleyEtAl2009,MassetCasoli2009,MassetCasoli2010,PaardekooperEtAl2010,PaardekooperEtAl2011}). In an isothermal disc, both the magnitude and sign 
of the corotation torque depend on the gradient of the surface density at the corotation radius.  


\subsection{Migration in magnetized discs}

\subsubsection{Migration in laminar MHD discs due to magnetic resonances}
\label{sec:migration in laminar MHD discs}

In strongly magnetized laminar discs, magnetic waves can be excited, and torques associated with the magnetic waves can affect a planet's migration. 
\citep[e.g.,][]{Terquem2003,FromangEtAl2005,FuLai2011}. \citet{Terquem2003} investigated the propagation of waves in a magnetized disc in which 
the magnetic field is purely azimuthal and found that there are two singular radii at which the frequency perturbation in a frame rotating with the fluid 
matches that of a slow MHD wave propagating along the field lines; these radii define the locations of the magnetic resonances. The inner and outer 
magnetic resonances are denoted similarly to Lindblad resonances, with the inner magnetic resonances $r_{\rm IMR} < r_{\rm p}$, and the outer 
magnetic resonances $r_{\rm OMR} > r_{\rm p}$. \citet{Terquem2003} derived the following dispersion relation:
\begin{equation}
	m^{2}(\Omega-\Omega_{\rm p})^{2} = \frac{m^{2}c_{s}^{2}v_{\rm A}^{2}}{r^{2}(v_{A}^{2}+c_{s}^{2})}.
	\label{eqn:mag-res-dispersion}
\end{equation}
Here, $v_{\rm A}$ is the Alfv\'{e}n speed, given by \citet{Terquem2003} as $ v_{\rm A}^{2} = {\langle B^{2}\rangle}/{4\pi\Sigma}$, where 
$\langle B^{2} \rangle=\int B^2 dz$ is the vertically-integrated square of the magnetic field. The sound speed 
$c_{s}^2=d\langle P\rangle/d\Sigma$, where $\langle P\rangle=\int{Pdz}$. \citet{Terquem2003} determines the strength of the magnetic field via
$\beta_{v_a} = {c_{s}^{2}}/{v_{A}^{2}}$, which is evaluated at the location of the planet.
\footnote{In our simulations, we use the standard definition of the plasma parameter, $\beta~=~2\beta_{v_A}$.}

The locations of the resonances when the disc is Keplerian ($\Omega \approx \Omega_{\rm K}$) and thin ($H/r \ll 1$) are given by
\begin{equation}
	|r_M-r_p|=\frac{2H}{3\sqrt{1+\beta_{v_A}}},
	\label{eqn:rm}
\end{equation}
where the thickness of the disc $H$ and the plasma parameter $\beta_{v_A}=c_s^2/v_A^2$ are evaluated at $r = r_{\rm p}$. As the field becomes 
weaker (i.e., $\beta_{v_A} \rightarrow \infty$), the magnetic resonances converge toward the corotation radius. We suggest that 
$c_s/(r\Omega)\approx H/r$ and find the position of the inner and outer magnetic resonances to be
\begin{equation}
    r_{\rm IMR} = r_{\rm p} - \frac{2H}{3\sqrt{1+\beta_{v_A}}},
    ~~~~
    r_{\rm OMR} = r_{\rm p} + \frac{2H}{3\sqrt{1+\beta_{v_A}}}.
    \label{terquem_rOMR}
\end{equation}
\citet{Terquem2003} showed that the waves associated with the magnetic resonances can exert a positive torque on the planet that is larger in 
magnitude than the differential Lindblad torque if the magnetic field increases toward the star.

\citet{FromangEtAl2005} performed simulations of a planet's migration in a magnetized disc and found that the magnetic resonances can slow or stop 
the migration of a planet. In particular, they investigated the migration of a $5 M_{\oplus}$ planet embedded in a disc that is threaded with an azimuthal 
magnetic field with $B_{\varphi}\propto r^{-k}$ with $\Sigma = {\rm const}$ initially. They took an initial value of 
$\beta_{v_a} = 2$ at the planet's location and found that, depending on the steepness of the magnetic field distribution (defined by $k$),  
the planet's migration slowed or reversed in direction.

\citet{GuiletEtAl2013} studied the effects of an MHD corotation torque on a 
low-mass planet in a 2D laminar disc with a weak azimuthal field threading the disc. The field was not strong enough to generate an appreciable torque 
from magnetic resonances, and it was not strong enough to dominate the hydrodynamic corotation torques, but a ``torque excess'' attributed to the 
presence of the magnetic field was found. 


\subsubsection{Migration in a turbulent disc}
\label{sec:condition for mri}

The MRI arises under conditions in which a weak magnetic field threads the disc and the radial component of the field can be stretched and enhanced
by the differential rotation in the disc. Such a seed field can either be an axial field, perpendicular to the disc
\citep{BalbusHawley1991,BalbusHawley1998}, or an azimuthal field \citep{TerquemPapaloizou1996}. Below, we briefly summarize the conditions for the
onset of the MRI instability in these two cases.

\smallskip

\textit{Axial Seed Magnetic Field.}

\noindent \citet{BalbusHawley1991} considered the case in which an axial magnetic field, $B_0\hat{\bf z}$, threads a
Keplerian disc that rotates with an angular speed $\Omega$. For axisymmetric perturbations of the disc, for which
$\delta{\bf v}= [\delta v_r(z,t), \delta v_\varphi(z,t),0]$ and $\delta {\bf B}~=~[\delta B_r(z,t),\delta
B_\varphi(z,t),0]$, and for perturbations proportional to $\exp(ik_z z - i\omega t)$, one finds the dispersion relation
\begin{equation}
    \omega_\pm^2 = (k_z v_A)^2 +{1\over 2}\kappa_r^2 \pm \left[{1\over 4}\kappa_r^4+4(k_z v_A\Omega)^2\right]^{1/2},
\end{equation}
 where $v_A \equiv B_0/\sqrt{4\pi \rho}$  is the Alfv\'en velocity, and
\begin{equation}
    \kappa_r \equiv [4\Omega^2 +2r\Omega d\Omega/dr]^{1/2}
\end{equation}
is the radial epicyclic frequency of the disc. In order for the perturbation to fit within the vertical extent of the disc,  one needs $k_z h \gtrsim 1$, where
$h = {c_s}/{\Omega}$ is the half-thickness of the disc and $c_s$ is the isothermal sound speed in the disc.  For most conditions, the disc is thin, with
$h \ll r$ or $c_s \ll r\Omega$.

Evidently, instability can occur if $\omega_-^2 <0$, which happens if $(k_{z}v_A)^2 < -r \Omega d\Omega^2/dr$.  For a
Keplerian disc, this corresponds to $(k_z v_A)^2  < 3\Omega^2$. Therefore, the above-mentioned condition that $k_z h
\gtrsim 1$ implies that instability occurs only for $ v_A < c_s$.

\smallskip

\textit{Toroidal seed magnetic field.}

\noindent \citet{TerquemPapaloizou1996} studied the linear MHD stability/instability of a thin Keplerian disc with a
toroidal magnetic field $B_\varphi(r,z)$. This case is more complicated than the case of a vertical field
\citep{BalbusHawley1998}.

The complication in the toroidal field case results from the presence of both the MRI instability and the buoyancy
instability of the toroidal field \citep{HoyleIreland1960,Parker1966}. The buoyancy instability is triggered by an
azimuthally-dependent, radially-localized vertical displacement in the plasma and the toroidal field
\citep{TerquemPapaloizou1996}. These authors find that the MRI instability in a thin disc with an embedded toroidal
magnetic field shows up in localized  perturbations (i.e., whose wavelengths are small compared with $r$), $\propto
\exp(i k_r r +i m \varphi + i k_z z - i \omega t)$, under the conditions $k_z^2 \gg k_r^2$ and $(k_\varphi v_A)^2  < - r
d\Omega^2/dr$, where $k_\varphi = m/r$.  This is the same as the condition for the MRI instability in a disc threaded
by a vertical field with $B_z \rightarrow B_\varphi$ and $k_z \rightarrow k_\varphi$. Therefore, perturbations of the
azimuthal field may also lead to MRI turbulence.

Work done by \citet{BaruteauEtAl2011} and \citet{UribeEtAl2011} showed the existence of additional MHD corotation
torques in MRI-turbulent discs using 3D MHD simulations. Furthermore, \citet{NelsonPapaloizou2004} investigated planet
migration in an MRI-turbulent disc and observed stochastic migration. However, the origin of the torque on the planet
in a turbulent disc is not well studied.

%
%

\section{Model}
\label{sec:model}

\subsection{MHD Equations}
\label{subsec:MHD equations}

We utilize the MHD equations to numerically evaluate the perturbative effect of the planet on the disc:
\begin{enumerate}
	\item Continuity equation (conservation of mass)
		\begin{equation}
			\frac{\partial\Sigma}{\partial t} 
                            + \frac{1}{r}\frac{\partial}{\partial r}(r\Sigma v_{r}) 
                            + \frac{1}{r}\frac{\partial}{\partial\varphi}(\Sigma v_{\varphi}) = 0,
		\label{eqn:continuity}
		\end{equation}
	where $\Sigma=\int\rho dz$ is the surface density (with $\rho$ the volume density), and $v_{r}$ and $v_{\varphi}$ are the radial and azimuthal 
	velocities, respectively.

	\item Radial equation of motion (conservation of momentum)
	\begin{eqnarray}
		\nonumber \frac{\partial}{\partial t}(\Sigma v_{r}) 
                                        &+& \frac{1}{r}\frac{\partial}{\partial r}\left[r\left(\Sigma v_{r}^{2} 
                                       + \Pi + \frac{\Psi_{rr}+\Psi_{\varphi\varphi}}{8\pi} - \frac{\Psi_{rr}}{4\pi}\right)\right] \\
		\nonumber &+& \frac{1}{r}\frac{\partial}{\partial\varphi}\left(\Sigma v_{r}v_{\varphi}-\frac{\Psi_{r\varphi}}{4\pi}\right) \\
                                        &=& \frac{\Pi}{r} +\frac{\Psi_{rr} + \Psi_{\varphi\varphi}}{8\pi r} - \Sigma\frac{GM_{\star}}{r^{2}} + \Sigma w_{r},
	\label{eqn:radial_motion}
	\end{eqnarray}
	where $\Pi = \int P dz$ is the surface pressure (with $P$ the volume pressure), $\Sigma w_{r}$ is the radial force exerted on the disc by the 
	planet (per unit area of the disc), and $\Psi_{rr}$, $\Psi_{r\varphi}$, and $\Psi_{\varphi\varphi}$ are magnetic surface variables that are
	discussed in more detail in \S \ref{subsec:mag_variables}.

	\item Azimuthal equation of motion (conservation of angular momentum)
		\begin{eqnarray}
			\nonumber \frac{\partial}{\partial t}(\Sigma v_{\varphi})
                                &+& \frac{1}{r^{2}}\frac{\partial}{\partial r}
			\left[r^{2}\left(\Sigma v_{r}v_{\varphi}-\frac{\Psi_{r\varphi}}{4\pi}\right)\right] \\     				\nonumber
                                 &+& \frac{1}{r}\frac{\partial}{\partial\varphi}
                                \left(\Sigma v_{\varphi}^{2} + \Pi + \frac{\Psi_{rr}+\Psi_{\varphi\varphi}}{8\pi} - \frac{\Psi_{\varphi\varphi}}{4\pi}\right) \\
                                 &=& \Sigma w_{\varphi},
			\label{eqn:az_motion}
		\end{eqnarray}
	where $\Sigma w_{\varphi}$ is the azimuthal force exerted on the disc by the planet (per unit area of the disc).

	\item Radial induction equation
		\begin{equation}
			\frac{\partial\Phi_{r}}{\partial t} 
                            + \frac{1}{r}\frac{\partial}{\partial\varphi}\left(v_{\varphi}\Phi_{r} - v_{r}\Phi_{\varphi}\right) = 0,
			\label{eqn:radial_induction}
		\end{equation}
	where $\Phi_{r}$ and $\Phi_{\varphi}$ are magnetic surface variables, also discussed in \S \ref{subsec:mag_variables}.

	\item Azimuthal induction equation
		\begin{equation}
			\frac{\partial\Phi_{\varphi}}{\partial t}+ \frac{\partial}{\partial r}\left(v_{r}\Phi_{\varphi} - v_{\varphi}\Phi_{r}\right) = 0.
			\label{eqn:az_induction}
		\end{equation}

	\item Entropy balance equation
		\begin{equation}
			\frac{\partial}{\partial t}\left(\Sigma S\right) 
		       + \frac{1}{r}\frac{\partial}{\partial r}\left(r\Sigma S v_{r}\right)
                            + \frac{1}{r}\frac{\partial}{\partial\varphi}(\Sigma S v_{\varphi}) = 0,
		\end{equation}
	where $S = {\Pi}/{\Sigma^{\gamma}}$ is a function analogous to entropy, and we use $\gamma = 1.01$ so that our disc is isothermal. We chose
	an isothermal disc to ease comparisons of our results with the results of other authors (e.g., by \citealt{FromangEtAl2005}).
\end{enumerate}

Moreover, viscosity terms were added to the equations of motion following the $\alpha$ prescription of \citet{ShakuraSunyaev1973} (see details in
\citealt{KoldobaEtAl2015}). We use a very small viscosity ($\alpha= 0.001$), analogous to \citet{FromangEtAl2005}, to isolate the effects of slow 
magnetosonic waves in the disc by smoothing out the effects of fast magnetosonic waves.


\subsubsection{Magnetic surface variables}\label{subsec:mag_variables}

The ``volume'' values for the radial and azimuthal magnetic fields are given by $B_{r}$ and $B_{\varphi}$. Their vertically-integrated counterparts may 
be defined similarly to $\Sigma$ and $\Pi$ as
\begin{align}
	\Phi_{r} = \int B_{r} dz &  &{\rm and} & & \Phi_{\varphi} = \int B_{\varphi} dz,
	\label{eqn:phi-def}
\end{align}
respectively. There are also terms in the MHD equations involving magnetic flux or energy that involve products of these variables: $B_{r}^{2}$, 
$B_{\varphi}^{2}$, and $B_{r}B_{\varphi}$. We define the following vertically-integrated quantities for these products,
\begin{align}
	\Psi_{rr}=\int B_{r}^{2} dz, &  & \Psi_{\varphi\varphi} = \int B_{\varphi}^{2} dz, & & {\rm and} & & \Psi_{r\varphi} = \int B_{r}B_{\varphi} dz,
	\label{eqn:psi-def}
\end{align}
respectively.

As shown in Eqns. (\ref{eqn:radial_motion}) - (\ref{eqn:az_induction}), the induction equations use $\Phi_{r}$ and $\Phi_{\varphi}$, while the 
equations of motion use $\Psi_{rr}$, $\Psi_{r\varphi}$, and $\Psi_{\varphi\varphi}$. As such, we need a way to relate $\Phi$ and $\Psi$. We can do 
this by introducing a ``magnetic'' thickness of the disc, $h_m$. Using the definitions of $\Phi$ and $\Psi$, with $h_m$, we find that
\begin{align}
	\Psi_{rr}                    &= \frac{\Phi_{r}\Phi_{r}}{h_m}; 
       & \Psi_{r\varphi}           &= \frac{\Phi_{r}\Phi_{\varphi}}{h_m}; 
       & \Psi_{\varphi\varphi} &= 	\frac{\Phi_{\varphi}\Phi_{\varphi}}{h_m}.
	\label{eqn:Psi_Phi}
\end{align}
We suggest that the value of $h_m = {\rm const}$ is the same in all three relations. By relating $\Phi$ and $\Psi$ in this way, we can define a 
``surface magnetic field'', $\mathfrak{B}$, such that the MHD equations are parameterized with respect to a single magnetic field variable,
\begin{align}
	\mathfrak{B}_{r} &= \frac{\Phi_{r}}{\sqrt{h_m}} &{\rm and} &  & \mathfrak{B}_{\varphi} &= \frac{\Phi_{\varphi}}{\sqrt{h_m}}.
	\label{eqn:F_Phi}
\end{align}
Then, the magnetic terms in Equations (\ref{eqn:radial_motion}) - (\ref{eqn:az_induction}) take their usual form.

The radial equation of motion becomes
\begin{eqnarray}
	\nonumber \frac{\partial}{\partial t}(\Sigma v_{r}) &
		        +& \frac{1}{r}\frac{\partial}{\partial r}\left[r\left(\Sigma v_{r}^{2} + \Pi
		        + \frac{\mathfrak{B}_{r}^{2}+\mathfrak{B}_{\varphi}^{2}}{8\pi} - \frac{\mathfrak{B}_{r}^{2}}{4\pi}\right)\right] \nonumber \\
	&+& \frac{1}{r}\frac{\partial}{\partial\varphi}
                   \left(\Sigma v_{r}v_{\varphi}-\frac{\mathfrak{B}_{r}\mathfrak{B}_{\varphi}}{4\pi}\right) \nonumber \\
	&=& \frac{\Sigma v_{\varphi}^{2}}{r} + \frac{\Pi}{r} +\frac{\mathfrak{B}_{r}^{2} + \mathfrak{B}_{\varphi}^{2}}{8\pi r} \nonumber \\
	&-& \Sigma\frac{GM_{\star}}{r^{2}} + \Sigma w_{r},
	\label{eqn:radial-eqn-motion-final}
\end{eqnarray}
and the azimuthal equation of motion becomes
\begin{eqnarray}
	\frac{\partial}{\partial t}(\Sigma v_{\varphi}) &+& \frac{1}{r^{2}}\frac{\partial}{\partial r}\left[r^{2}
	\left(\Sigma v_{r}v_{\varphi}-\frac{\mathfrak{B}_{r}\mathfrak{B}_{\varphi}}{4\pi}\right)\right] \nonumber \\
	&+& \frac{1}{r}\frac{\partial}{\partial\varphi} \left(\Sigma v_{\varphi}^{2} + \Pi 
								     + \frac{\mathfrak{B}_{r}^{2} + \mathfrak{B}_{\varphi}^{2}}{8\pi} 
								      - \frac{\mathfrak{B}_{\varphi}^{2}}{4\pi}\right) \nonumber \\
	&=& \Sigma w_{\varphi}.
	\label{eqn:az-eqn-motion-final}
\end{eqnarray}
The radial and azimuthal induction equations become, respectively,
\begin{equation}
	\frac{\partial \mathfrak{B}_{r}}{\partial t} + \frac{1}{r}\frac{\partial}{\partial\varphi}
	\left(v_{\varphi}\mathfrak{B}_{r} - v_{r}\mathfrak{B}_{\varphi}\right) = 0
	\label{eqn:radial-ind-final}
\end{equation}
and
\begin{equation}
	\frac{\partial \mathfrak{B}_{\varphi}}{\partial t}+ \frac{\partial}{\partial r}
	\left(v_{r}\mathfrak{B}_{\varphi} - v_{\varphi}\mathfrak{B}_{r}\right) = 0.
	\label{eqn:az-ind-final}
\end{equation}

\subsection{Planetary equation of motion}
\label{subsec:equation of motion}

We calculate the equations of motion in the stellar reference frame, which is not inertial because the star also revolves about the center of mass of the 
system. So, an inertial force term is added to the equation of motion for both the disc and the planet. Assuming that the inertial acceleration is only due 
to the gravitational attraction between the star and the planet (but not the disc), the inertial force per unit mass (i.e., acceleration) is
\begin{equation}
	\mathbf{w}_{i} = -\frac{GM_{\rm p}}{r_{\rm p}^{3}}\mathbf{r}_{\rm p}.
\end{equation}
To describe the gravitational influence of the planet on the disc, we use a gravitational potential similar to that  used by \citet{FromangEtAl2005},
\begin{equation}
	\Phi_{\rm p} =  - \frac{GM_{\rm p}}{\sqrt{r^{2} +r_{\rm p}^{2} - 2rr_{\rm p}\cos(\varphi-\varphi_{\rm p}) + \epsilon^2}},
	\label{eqn:grav-potential}
\end{equation}
where $\epsilon = 0.1H$ is the gravitational smoothing length, and $H$ is the scale height of the disc. The total force per unit mass is
\begin{equation}
	\mathbf{w} = \mathbf{w}_{p} + \mathbf{w}_{i}, ~~~ \mathbf{w}_{p}=- \nabla{\Phi_{\rm p}}.
	\label{eqn:tot-force-per-unit-mass}
\end{equation}
The components of this force in polar coordinates, $w_{r}$ and $w_{\varphi}$, are used in Eqs. (\ref{eqn:radial-eqn-motion-final}) and 
(\ref{eqn:az-eqn-motion-final}).

The force exerted {\it on} the planet by a particular fluid element with mass $dM = \Sigma r dr d\varphi$, is the acceleration given in 
Eqn. (\ref{eqn:tot-force-per-unit-mass}), with opposite sign, multiplied by the mass of the fluid element,
\begin{equation}
	d\mathbf{f}_{\rm disc\rightarrow p} = -dM\mathbf{w}_p = dM\nabla\Phi_{\rm p}.
	\label{eqn:df}
\end{equation}
We then calculate the total force exerted on the planet by the disc by integrating over the disc within the computational domain,
\begin{equation}
	\mathbf{F}_{\rm disc\rightarrow p} = \int_{\rm disc} d\mathbf{f}_{\rm disc\rightarrow p} = \int_{\rm disc}  dM \nabla\Phi_{\rm p},
	\label{eqn:planet-eqn-of-motion}
\end{equation}
which we use to find the position ($\mathbf{r}_{\rm p}$) and velocity ($\mathbf{v}_{\rm p}$) of the planet at each time step via the planet's equation 
of motion:
\begin{equation}
	M_{p} \frac{d\mathbf{v}_{\rm p}}{dt} = -\frac{GM_{\star}M_{\rm p}}{r_{\rm p}^{3}}\mathbf{r}_{\rm p}
                                                                              -\frac{GM_{\rm p}^{2}}{r_{\rm p}^{3}}\mathbf{r}_{\rm p} + \mathbf{F}_{\rm disc\rightarrow p}.
\end{equation}
The gravitational torque in the $z$ direction on the planet is the sum over the torques from each fluid element:
\begin{equation}
	T_{z} = \int_{\rm disc} \left[\mathbf{r} \times  d\mathbf{f}_{\rm disc\rightarrow p} \right]_{z}.
\end{equation}
We also calculate the planet's orbital energy and angular momentum per unit mass (e.g., \citealt{MurrayDermott1999}) via
\begin{align}
	E = \frac{1}{2}\lvert{\mathbf v}_{\rm p}\rvert^{2} - \frac{GM_{\star}}{r_p} 
	& & {\rm and} 
	& &  L = \mathbf{r}_{\rm p} \times \mathbf{v}_{\rm p},
\end{align}
respectively. We use these relationships to calculate the semimajor axis and eccentricity of the planet's orbit at each time step,
\begin{align}
	a = -\frac{1}{2}\frac{GM_{\star}}{E} & & {\rm and} 
	& & 
	e = \sqrt{1 - \frac{L^{2}}{GM_{\star}a}},
\end{align}
respectively.


\subsection{Initial conditions}\label{num-model:ic-bc}

The initial conditions for the surface density, surface pressure and surface ``magnetic field" are defined to be power laws
\begin{align}
	&\Sigma = \Sigma_{i} \left( \frac{r}{r_{i}}\right)^{-n}, \\
	&\Pi = \Pi_{i} \left( \frac{r}{r_{i}}\right)^{-l}, ~~~~{\rm and}\\
	&\mathfrak{B}_{\varphi} = \mathfrak{B}_{\varphi,i} \left(\frac{r}{r_{i}}\right)^{-k},
	\label{eqn:init-power-laws}
\end{align}
where ${r}_{i}$ is a characteristic radius in the disc; in most of our simulations $r_{i} = r_{{\rm p},i}$ is the initial location of the planet. Additionally, 
$n$, $l$, and $k$ are the power laws in the density, pressure, and magnetic field distributions, respectively. We suggest that the initial pressure 
distribution is similar to the initial density distribution (i.e., $n=l$). The initial sound speed in the disc at ${r}_{i}$, defined as a fraction of the 
Keplerian speed at this radius, $k_{s}$, is
\begin{equation}
	c_{i}^{2}=k_{s} \frac{GM_{\star}}{{r}_{i}},
	\label{eqn:init-c2}
\end{equation}
where $k_{s} = 0.01$. The initial surface pressure at ${r}_{i}$ is $\Pi_{i}~=~\Sigma_{i} c_{i}^{2}$. The initial plasma parameter at ${r}_{i}$ is then 
defined as
\begin{equation}
	\beta_{i}=\frac{8\pi\Pi_{i}}{\mathfrak{B}^{2}_{\varphi,i}}.
	\label{eqn:init-beta}
\end{equation}
Note again that this plasma parameter, $\beta_{i}$, is twice as large as the plasma parameter defined in \citet{FromangEtAl2005}. We determine the 
surface magnetic field distribution via
\begin{equation}
    \mathfrak{B}_{\varphi,i}=\sqrt{\frac{8 \pi \Pi_{i}}{\beta_{i}}}.
\end{equation}

We determine the initial equilibrium in the disc from the force balance in the radial direction:
\begin{equation}
	- \frac{v_\varphi^2}{r} + \frac{1}{\Sigma} \prodi{\Pi}{r} + \frac{1}{8 \pi r^2 \Sigma} \prodi{(r \mathfrak{B}_\varphi)^2}{r} 
	+ \frac{G M_{\star}}{r^2} = 0. 
	\label{eqn:init-force-balance-disc}
\end{equation}
It is satisfied if the initial distribution of the azimuthal velocity has the form
\begin{equation}
	v_{\varphi}^{2} = \frac{GM_{\star}}{r} + c^{2}_{i} \left( \frac{r}{r_{i}} \right)^n \bigg[ -n \left( \frac{r_{i}}{r} \right)^n
			      + \frac{1-k}{\beta_{i}} \left( \frac{r_{i}}{r} \right)^{2k} \bigg].
	\label{eqn:init-az-vel-dist}
\end{equation}

For all of the simulations presented here, we took $k_{s}~=~0.01$. The thickness of the disc at 
${r}_{i}$ is $H_{i}~=~\sqrt{k_s {r}_{i}}$. The smoothing radius of the gravitational potential is $\epsilon~=~0.1H$. An initial plasma 
parameter of $\beta_{i} = 1, 2$ is taken for analysis of migration in laminar MHD discs, while we increase $\beta_{i}$  up to $\beta_{i}=100$ to study 
migration in a turbulent disc. We consider the migration of a $5 M_{\oplus}$ planet in most of the simulations, as well as a $20 M_{\oplus}$ planet in a 
few simulations.

\begin{table}
\begin{tabular}{l l l}
\hline
\multicolumn{3}{l}{\textbf{Variable Definitions}}\\
\hline
$r_{0}$ & \multicolumn{2}{l}{Reference distance} \\
$v_{0}$ & \multicolumn{2}{l}{Reference velocity} \\
$P_{0}$ & \multicolumn{2}{l}{Reference orbital period} \\
$M_{{\rm d}0}$ & \multicolumn{2}{l}{Reference disc mass} \\
$\Sigma_{0}$ & \multicolumn{2}{l}{Reference surface density} \\
$B_{0}$ & \multicolumn{2}{l}{Reference magnetic field strength} \\
$T_{0}$ & \multicolumn{2}{l}{Reference torque per unit mass} \\
\hline
\multicolumn{3}{c}{(a)}\\ \\
\hline
\multicolumn{3}{l}{\textbf{Standard reference values}}\\
\hline
 & $r_{0} = 0.1$ AU & $r_{0} = 1$ AU \\
\hline
$v_{0}$ & $94.1$ km s$^{-1}$ & $29.8$ km s$^{-1}$\\
$P_{0}$ & $11.6$ days & $367$ days\\
$M_{{\rm d}0}$ & $1 M_{\odot}$& $1 M_{\odot}$\\
$\Sigma_{0}$ & $8.84\times 10^{8}$ g cm$^{-2}$ & $8.84\times 10^{6}$ g cm$^{-2}$\\
$B_{0}$ & $228$ kG & $2.28$ kG\\
$T_{0}$ & $8.85\times 10^{13}$ cm$^{2}$ s$^{-2}$  & $8.85\times 10^{12}$ cm$^{2}$ s$^{-2}$\\
\hline
\multicolumn{3}{c}{(b)}\\ \\
\hline
\multicolumn{3}{l}{\textbf{Rescaled reference values}}\\
\hline
 & $r_{0} = 0.1$ AU & $r_{0} = 1$ AU \\
\hline
$v_{0}$ & $94.1$ km s$^{-1}$ & $29.8$ km s$^{-1}$  \\
$P_{0}$ & $11.6$ days & $367$ days\\
$M_{{\rm d}0}$ & $10^{-3} M_{\odot}$ & $10^{-3} M_{\odot}$\\
$\Sigma_{0}$ & $8.84\times 10^{5}$ g cm$^{-2}$ & $8.84\times 10^{3}$ g cm$^{-2}$\\
$B_{0}$ & $7.22$ kG & $72.2$ G\\
$T_{0}$ & $8.85\times 10^{13}$ cm$^{2}$ s$^{-2}$ & $8.85\times 10^{12}$ cm$^{2}$ s$^{-2}$ \\
\hline
\multicolumn{3}{c}{(c)}
\end{tabular}
\caption{Example reference units calculation. The stellar mass is $M_{\star} = 1 M_{\odot}$.
(a) Definitions of the variables used. (b) Standard reference values, corresponding to
$\widetilde{\Sigma} = 1$. (c) Rescaled reference values, corresponding to $\widetilde{\Sigma} = 0.001$,
which is the value used in the code.}
\label{table:units}
\end{table}


\subsection{Reference Units}\label{sec:reference units}

Our simulations are performed using dimensionless units $\widetilde{A}~=~A/A_{0}$ where $A_{0}$ are the reference units. We first choose some 
reference distance, $r_{0}$. The results of our simulations are applicable to multiple regions in a protoplanetary disc, because $r_{0}$ can be chosen to 
correspond to different regions of the disc. In Table \ref{table:units}, we show examples of reference values for scale distances of $r_{0} = 0.1$ AU 
and $r_{0} = 1$ AU. The thickness of the disc is $H~=~0.1 r_{0}$. We take the mass of the star,  $M_{\star}=1 M_{\odot}$, and determine the 
reference velocity, which is the Keplerian velocity at $r_{0}$, $v_{0}=\sqrt{GM_\star/r_{0}}$. The reference time is $t_{0}=r_{0}/v_{0}$, and we use 
the reference orbital period $P_{0}=2\pi t_{0}$ as the unit of time in our plots.

We next define the reference mass of the disc $M_{d0}$:  $M_{{\rm d}0}=M_{\star}$. 
We then introduce the reference surface density, $\Sigma_{0}$, such that $M_{{\rm d}0}=\Sigma_{0} r_{0}^{2}$. 
When the disc is homogeneous (i.e., $\Sigma = {\rm const}$), $M_{{\rm d}0}$ is the mass of the disc inside $r=r_{0}$. The reference surface pressure 
is $\Pi_{0}=\Sigma_{0} v_{0}^2$.

The reference surface magnetic field is derived from the condition $\Sigma_0 v_0^2=\mathfrak{B}_{0}^2$. Hence, 
$\mathfrak{B}_{0}=\sqrt{\Sigma_0 v_0^2}$. Taking into account our definitions of the surface magnetic field (see Eqns. \ref{eqn:phi-def} and 
\ref{eqn:psi-def}), we obtain the reference volume magnetic field: $B_{0}=\mathfrak{B}_{0}/\sqrt{r_{0}}$. The reference torque per unit mass is 
defined as $T_{0} = r_{0}^{2} / t_{0}^{2}$.
These reference units are used to convert equations from dimensionless units. We show examples in Table \ref{table:units}b.

In our model, we use a small value of the dimensionless density $\widetilde{\Sigma}=0.001$; as a result, our dimensional characteristic disc mass is 
$M_{d0} = 10^{-3} M_{\odot}$. The other characteristic values are closer to realistic values and are much smaller than the reference values shown in 
Table \ref{table:units}b. We show in Table \ref{table:units}c the typical dimensional 
values corresponding to our simulations. One can see that, in the case of $r_0=1$ AU, the surface density and other parameters are close to
those expected in real protoplanetary discs. The values shown for $r_0=0.1$AU are too large, but we keep these model parameters for all distances in 
order to compare our results with those in \citet{FromangEtAl2005} and others.\footnote{It is often the case that the disc mass and the surface 
density are taken to be larger than in realistic discs. This leads to more rapid migration and thus shorter computational times (see, e.g., 
\citealt{ArmitageRice2006}).} From this point forward, we will use only dimensionless units and remove tildes from variables.


\subsection{Grid and boundary conditions}\label{sec:grid and boundary}

We use a polar grid that is uniform in the $\varphi$ direction. In the $r$ direction, however, the grid is non-uniform; the size of the grid cells increases 
with radius such that the sides are approximately square-shaped throughout the disc. $N_{r}$ is defined to be the number of radial grid cells, and 
$N_{\varphi}$ is the number of azimuthal grid cells. Our grid consists of $N_{r} = 480$ and $N_{\varphi} = 1200$ cells. Our inner boundary is at 
$r_{\rm in} = 0.4 {r}_{{\rm p},i}$, and our outer boundary is at $r_{\rm out} = 5 {r}_{{\rm p},i}$.

We apply a wave damping procedure near both boundaries similar to that used in \S 3.1.3 of \citet{FromangEtAl2005} to avoid high-amplitude wave 
reflections that can overwhelm the migration signal. We perform damping after every time step for $r < r_{\rm damp,in}$, and $r > r_{\rm damp,out}$, 
where $r_{\rm damp,in}~=~1.375 r_{\rm in}$ and $r_{\rm damp,out} = 0.8 r_{\rm out}$. We calculate the velocities ($v_{r}$, $v_{\varphi}$), as well 
as $\Sigma$, $\Pi$, and $S$, using a method similar to \citet{FromangEtAl2005},
\begin{equation}
	\vec{\mathcal{J}} =
		\begin{cases}
			\mathcal{J}_{\rm in} + (\mathcal{J} - \mathcal{J}_{\rm in}) 
			\exp\left[ -\left(\frac{r-r_{\rm damp,in}}{\delta_{\rm in}}\right)^{2}\right] &  r < r_{\rm damp,in} \\
			\mathcal{J}_{\rm out} + (\mathcal{J} - \mathcal{J}_{\rm out}) 
			\exp\left[ -\left(\frac{r-r_{\rm damp,out}}{\delta_{\rm out}}\right)^{2}\right] &  r > r_{\rm damp,out} \\
			\mathcal{J} & {\rm otherwise}, \\
		\end{cases}
	\end{equation}
where $\vec{\mathcal{J}} = \left( v_{r},v_{\varphi},\Sigma,\Pi,S \right)$, $\delta_{\rm in} = 0.875 r_{\rm in}$, and 
$\delta_{\rm out}~=~0.8 r_{\rm out}$. Furthermore, $\mathcal{J}_{\rm in}$ and $\mathcal{J}_{\rm out}$ are the values of 
$\mathcal{J}$ at the inner and outer disc boundaries, respectively.

%
%
\section{Parameter space}\label{sec: parameter space}

We calculated a number of models at different initial parameters, which are described below (see also Table \ref{table:simulation_names}):
\newline

\noindent{\bf Simulation type:} The simulations are performed in a hydrodynamic disc or an MHD disc, or to study the interaction between the planet 
and waves in the disc.
\newline

\noindent{\bf Planet mass:} We explored two different planet masses -- $5 M_{\oplus}$ and $20 M_{\oplus}$.
\newline

\noindent{\bf Surface density exponent, $n$:} This defines the initial surface density distribution in the disc, according to $\Sigma \propto r^{-n}$. 
\newline

\noindent{\bf Surface magnetic field exponent, $k$:} This defines the initial surface magnetic field distribution in the disc, according to 
$\mathfrak{B} \propto r^{-k}$. 
\newline

\noindent{\bf Matter-to-magnetic pressure ratio, $\beta_{i}$:} This defines the value of $\beta_{i}$ at the initial location of the planet. 
\newline

\noindent{\bf Initial planet location, $r_{{\rm p},i}$:} This defines the initial orbital radius of the planet. \newline

\noindent The last column of Table \ref{table:simulation_names} shows the names of the simulations presented in subsequent sections. These names 
are referenced in the figure captions and related discussion.

\begin{table}
\begin{tabular} {l l l l l l l}
\hline
         & $M_{\rm p}$ &        &       &               &                                                       &  \\
Type &  ($M_{\oplus}$)      & $n$ & $k$ & $\beta_i$ & ${r}_{{\rm p},i}$ & Name        \\
\hline
Hydro  &  $5$ &  $-1$ & $-$ & $-$ & $1$ & H5n-1r1 \\
Hydro  &  $5$ &  $-0.5$ & $-$ & $-$ & $1$ & H5n-0.5r1 \\
Hydro  &  $5$ &  $0$ & $-$ & $-$ & $1$ & H5n0r1 \\
Hydro  &  $5$ &  $1$ & $-$ & $-$ & $1$ & H5n1r1 \\
Hydro  &  $20$ &  $-1$ & $-$ & $-$ & $1$ & H20n-1r1 \\
Hydro  &  $20$ &  $-0.5$ & $-$ & $-$ & $1$ & H20n-0.5r1 \\
Hydro  &  $20$ &  $0$ & $-$ & $-$ & $1$ & H20n0r1 \\
\hline
MHD  &  $5$ &  $0$ & $0$ & $1$ & $1$ & M5n0k0$\beta$1r1 \\
MHD  &  $5$ &  $0$ & $0$ & $2$ & $1$ & M5n0k0$\beta$2r1 \\
MHD  &  $5$ &  $0$ & $1$ & $2$ & $1$ & M5n0k1$\beta$2r1 \\
MHD  &  $5$ &  $0$ & $2$ & $2$ & $1$ & M5n0k2$\beta$2r1 \\
MHD  &  $5$ &  $0$ & $0$ & $10$ & $1$ & M5n0k0$\beta$10r1 \\
MHD  &  $5$ &  $0$ & $0$ & $100$ & $1$ & M5n0k0$\beta$100r1 \\
\hline
MHD  &  $5$ &  $1$ & $0$ & $2$ & $1$ & M5n1k0$\beta$2r1 \\
MHD  &  $5$ &  $1$ & $1$ & $1$ & $1$ & M5n1k1$\beta$1r1 \\
MHD  &  $5$ &  $1$ & $1$ & $2$ & $1$ & M5n1k1$\beta$2r1 \\
MHD  &  $5$ &  $1$ & $1$ & $10$ & $1$ & M5n1k1$\beta$10r1 \\
MHD  &  $5$ &  $1$ & $2$ & $2$ & $1$ & M5n1k2$\beta$2r1 \\
\hline
MHD  &  $20$ &  $0$ & $0$ & $2$ & $1$ & M20n0k0$\beta$2r1 \\
MHD  &  $20$ &  $0$ & $1$ & $2$ & $1$ & M20n0k1$\beta$2r1 \\
MHD  &  $20$ &  $0$ & $2$ & $2$ & $1$ & M20n0k2$\beta$2r1 \\
\hline
Waves & $5$ & $0$ & $-$ & $-$ & $2$ & W5n0k0r2 \\
Waves & $5$ & $0$ & $0$ & $100$ & $2$ & W5n0k0$\beta$100r2 \\
\hline
\end{tabular}
\caption{Simulation names and their respective distinguishing variables. See a more detailed description in \S \ref{sec: parameter space}.}
\label{table:simulation_names}
\end{table}

%
%

\section{Migration in a hydrodynamic disc}
\label{sec:migration hydro}

\begin{figure}
    \centering
    \includegraphics[width=7cm]{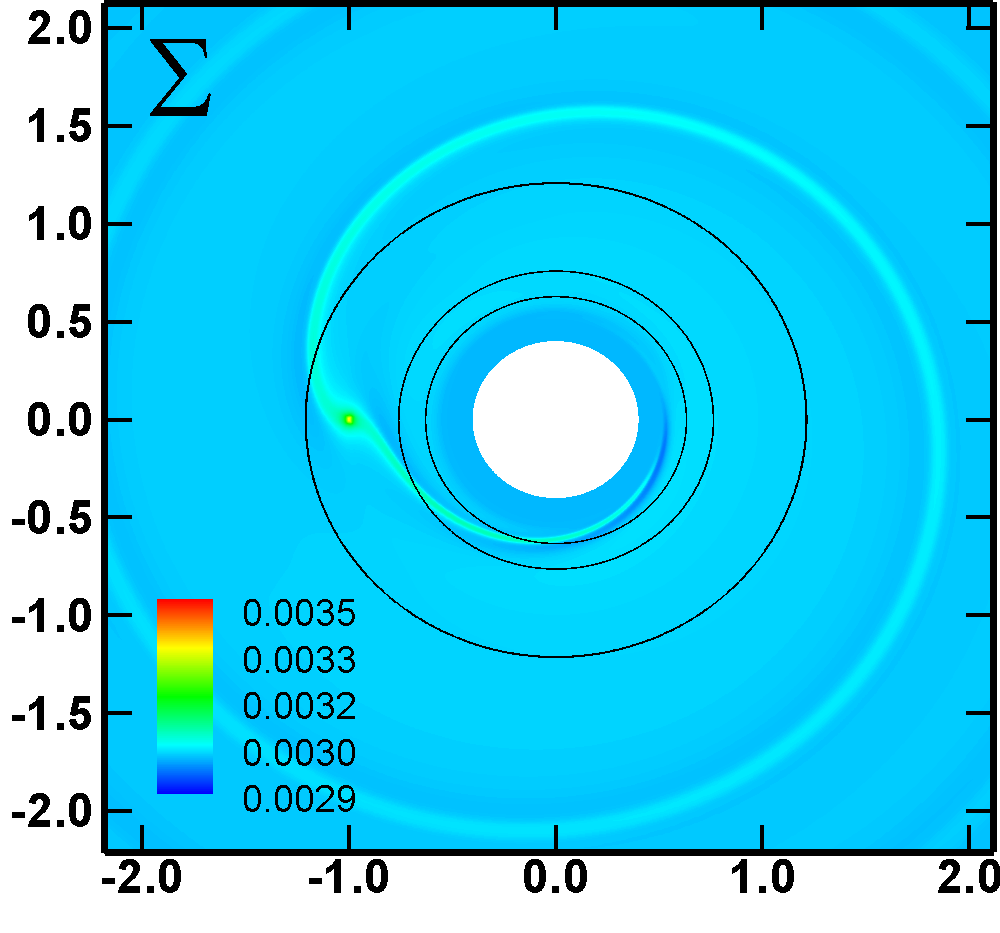}
    \caption{An example of a hydrodynamic simulation of migration corresponding to the model
                   H5n0r1 after $10$ orbits of the planet. The color background shows the surface density distribution.
	         The $m=1,2$ Lindblad resonances are indicated by the solid black circles.}
    \label{figure:hydro-sigma}
\end{figure}

\begin{figure*}
    \centering
    \includegraphics[width=6cm]{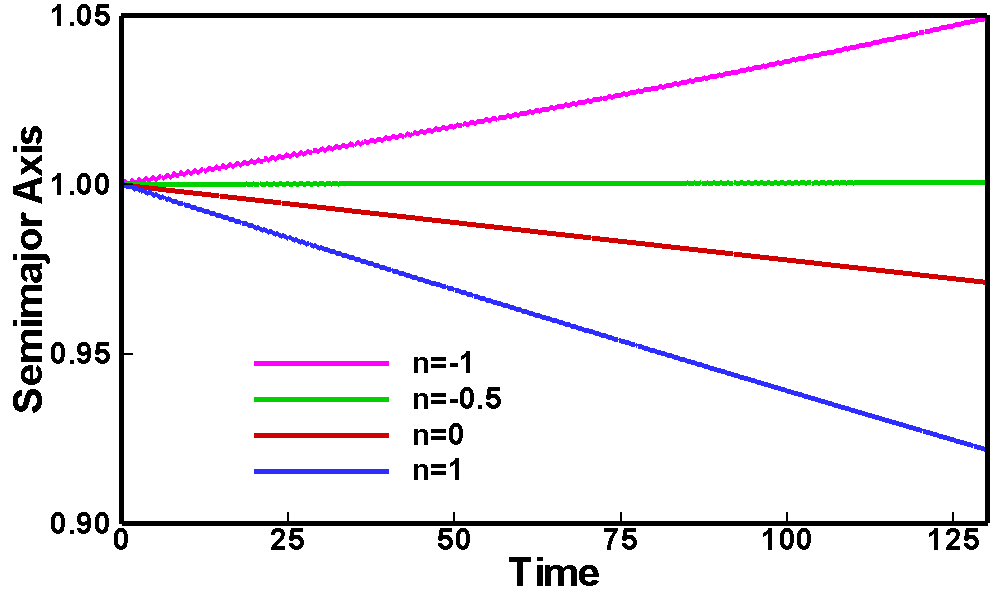}
    \includegraphics[width=6cm]{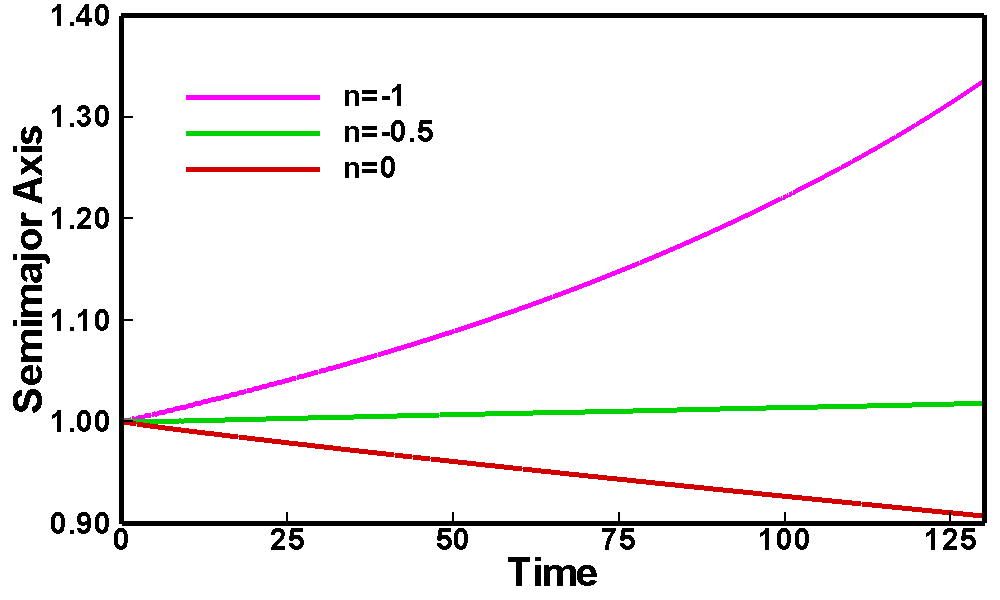}
    \caption{\textit{Left Panel:} The variation in semimajor axis versus time for a $5 M_{\oplus}$ 
     planet embedded in a hydrodynamic disc for different values of $n$ (where $\Sigma \propto r^{-n}$). 
     The models shown include: H5n-1r1, H5n-0.5r1, H5n0r1, H5n1r1.
     \textit{Right Panel:} A similar plot, but for a $20 M_{\oplus}$ planet. The models shown include: 
      H20n-1r1, H20n-0.5r1, H20n0r1.}
    \label{figure:hydro-5ME-20ME}
\end{figure*}

As a first step, we investigated the migration of a planet in a hydrodynamic disc with an initially homogeneous density distribution ($n=0$); these 
simulations are used as a base for studying migration in magnetic discs. Fig. \ref{figure:hydro-sigma} shows the surface density distribution in the disc 
after $10$ orbits of the planet. The planet excites two density waves at the inner and outer Lindblad resonances (which are shown as black circles in 
the figure).

As a next step, we considered different initial density distributions in the disc, $\Sigma\sim r^{-n}$ including those where the density is homogeneous, 
$n=0$, increases towards the star, $n=1$, or decreases toward the star, $n=-0.5, -1$. Fig. \ref{figure:hydro-5ME-20ME} (left panel) shows the 
variation of the planet's semimajor axis over time for these cases. We observed inward migration when $n=0$ and more rapid inward migration when
$n=1$. When $n=-0.5$, the migration almost stalls (where only slow outward migration has been observed), and we observed more rapid outward 
migration when $n=-1$.

The migration rate and its direction are determined by the cumulative value of the Lindblad and corotation torques, as described in Sec. 
\ref{section:hydro_torque}. When the density increases towards the star ($n=1$) or the density is constant in the disc ($n=0$), the Lindblad torque is 
larger than the corotation torque, and the planet migrates towards the star. However, when the density {\it decreases} towards the star, $n=-1$, the 
corotation torque is larger than the Lindblad torque, and the planet migrates outwards. In the case of a more shallow positive density distribution, 
$n=-0.5$, the Lindblad and corotation torques are almost equal, and the cumulative torque is small. 

The migration due to both the Lindblad and corotation torques has been studied in semi-analytical models by \citet{TanakaEtAl2002}. In two dimensions, 
they obtained the torque on the planet from the disc: 
\begin{equation}
	\Gamma_{\rm total,2D} = -(1.160+2.828\alpha)
                                                       \left( \frac{M_{\rm p}}{M_{\star}} \frac{r_{\rm p}\Omega_{\rm p}}{c_{s}}\right)^{2}
                                                       \Sigma(r_{\rm p})r_{\rm p}^{4}\Omega_{\rm p}^{2},
	\label{eqn:gamma-total}
\end{equation}
where $\alpha$ defines the slope of the density distribution ($\Sigma~\propto~r^{-\alpha}$), and $c_{s}$ is the isothermal sound speed in the disc at 
the orbital radius of the planet. This implies that the total torque on the planet in a two-dimensional disc is zero when $\alpha = -0.41$, suggesting that
\begin{enumerate}
	\item the torque on the planet is positive, and the planet migrates outward, when $\alpha < -0.41$;
	\item the torque on the planet is zero, and the planet's migration halts, when $\alpha = -0.41$; and
	\item the torque on the planet is negative, and the planet migrates inward, when $\alpha > -0.41$.
\end{enumerate}
The value of $n=-0.5$ obtained from our simulations is very close to the value $n=-0.41$ corresponding to zero torque derived by  
\citet{TanakaEtAl2002} for two dimensions.

We also performed simulations of the migration of a $20 M_{\oplus}$ planet. Fig. \ref{figure:hydro-5ME-20ME} (right panel) shows that the planet 
migrates inward when $n=0$; almost no migration is observed when $n=-0.5$ (as with the $5 M_{\oplus}$ planet); and the migration direction is 
outward when $n=-1$. These results are in accord with that of a $5 M_{\oplus}$ planet, while the migration rate is faster in the $20 M_{\oplus}$ case. 
This is expected according to Eqn. (\ref{eqn:gamma-total}), which states that the total Lindblad torque increases with the mass of the planet.

Our simulations of hydrodynamic discs are in accord with theoretical studies (e.g., \citealt{TanakaEtAl2002}) and simulations performed by others (see 
review by \citealt{KleyNelson2012}).

%
%

\section{Migration in laminar disc due to Magnetic Resonances}
\label{sec:magres migration}

\subsection{Magnetic resonances in the case of a constant density distribution}

\begin{figure*}
    \centering
    \includegraphics[width=11.cm]{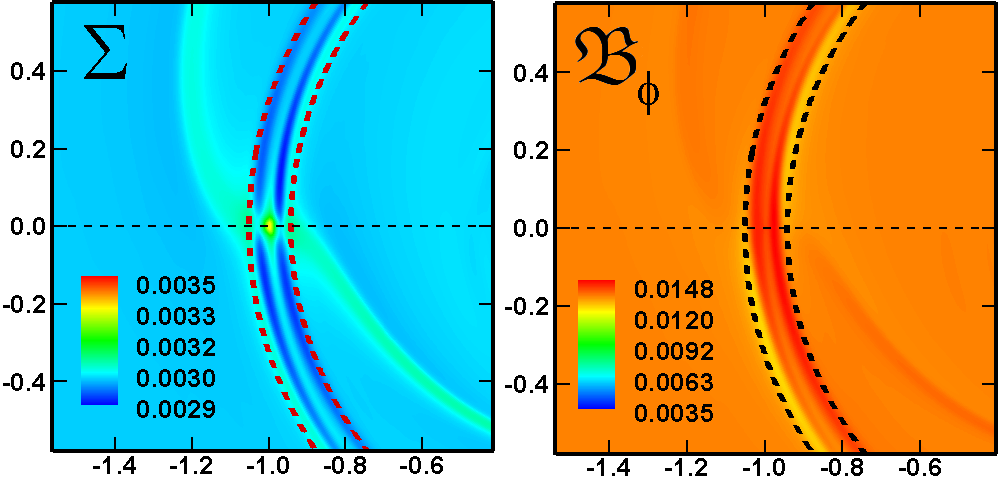}
    \includegraphics[width=5.0cm]{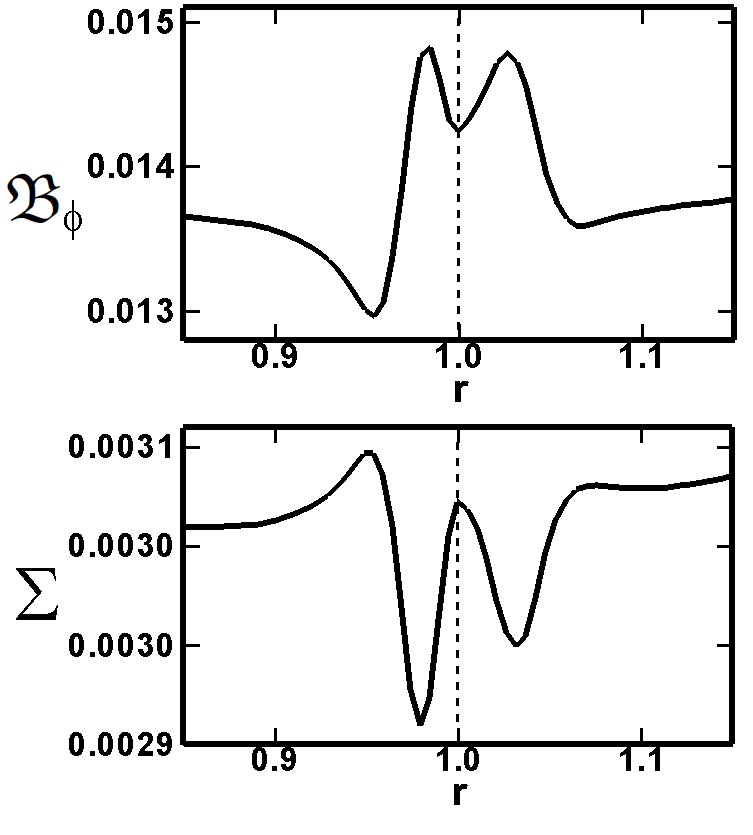}
    \caption{The surface density and surface azimuthal magnetic field distributions for the model M5n0k0$\beta$2r1 at $t=10$.
                  \textit{Left Panel:} The two-dimensional surface density distribution. The dashed red lines show the positions of the magnetic resonances.
                  \textit{Middle Panel:} The two-dimensional surface azimuthal magnetic field distribution. The dashed black lines show the positions of the magnetic resonances.
	       \textit{Right Top Panel:} The one-dimensional surface azimuthal magnetic field distribution, taken along the horizontal dashed line in the left two panels.}
                  \textit{Right Bottom Panel:} The one-dimensional surface density distribution, taken along the horizontal dashed line in the left two panels. The vertical dashed line shows the location
                                                          of the planet
    \label{figure:sigma-bf}
\end{figure*}

We performed a series of simulations to investigate the influence of an ordered azimuthal magnetic field in the disc on a planet's migration. We started by
investigating migration in discs with a homogeneous surface density distribution (i.e., $n=0$), with initial and boundary conditions very similar to 
those used by \citet{FromangEtAl2005}. In particular, we considered an isothermal disc (i.e., $\gamma=1.01$) and set the strength of the magnetic 
field near the planet such that $\beta_{i}=2$. Our goal was to see whether our code can reproduce the results presented by \citet{FromangEtAl2005}.

First, we investigated a homogeneous magnetic field (i.e., $k=0$). Fig. \ref{figure:sigma-bf} shows an example simulation after 10 orbital periods of 
the planet. We observed that the planet excites ring-like waves at which the azimuthal magnetic field is stronger and the surface density is lower than in 
other nearby regions in the disc. These locations correspond to the magnetic resonances that are excited by slow magnetosonic waves propagating 
along the field lines. Fig. \ref{figure:sigma-bf} shows two rings of lower density (left panel), corresponding to two rings of enhanced azimuthal field 
(middle panel). The position of the magnetic resonances is similar to that found in the simulations by \citet{FromangEtAl2005}, and they correspond to 
the theoretical resonance locations predicted by \citet{Terquem2003} (see the dashed-line circles in Fig. \ref{figure:sigma-bf}). The right panel shows 
the linear distribution of $B_{\varphi}$ and $\Sigma$ in the radial direction.

\begin{figure*}
    \centering
    \includegraphics[width=5.855cm]{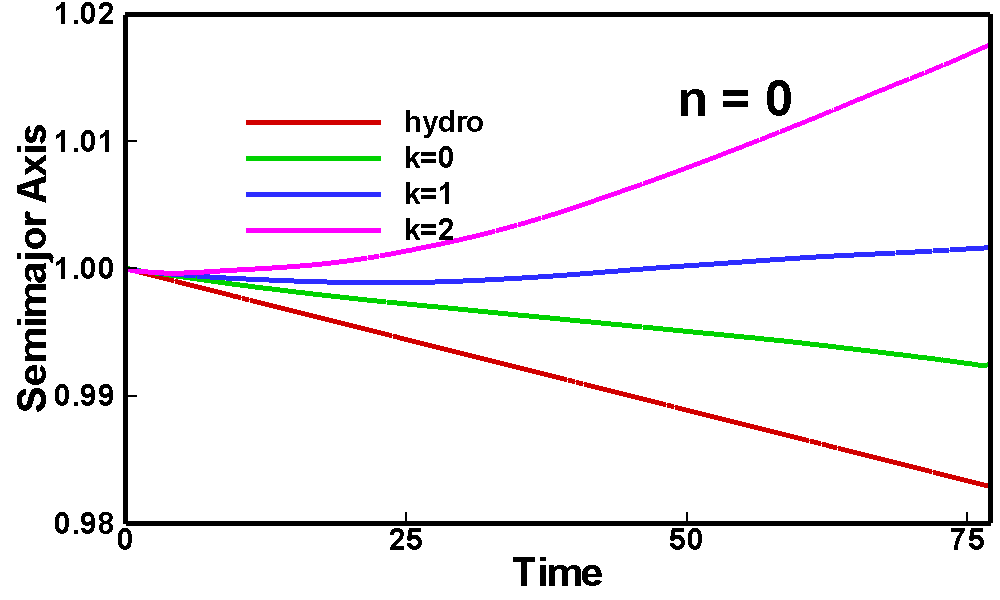}
    \includegraphics[width=5.855cm]{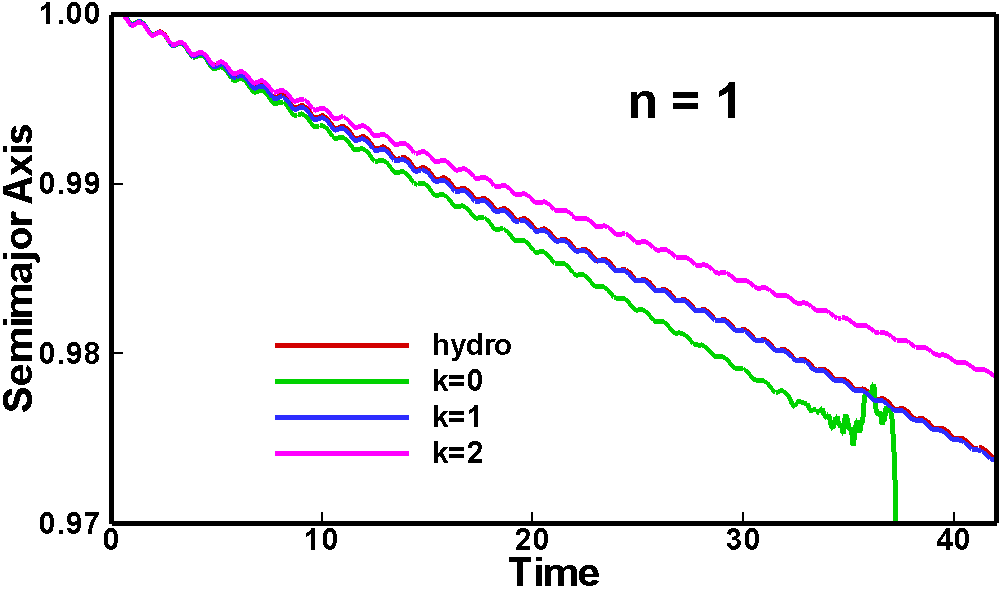}
    \includegraphics[width=5.855cm]{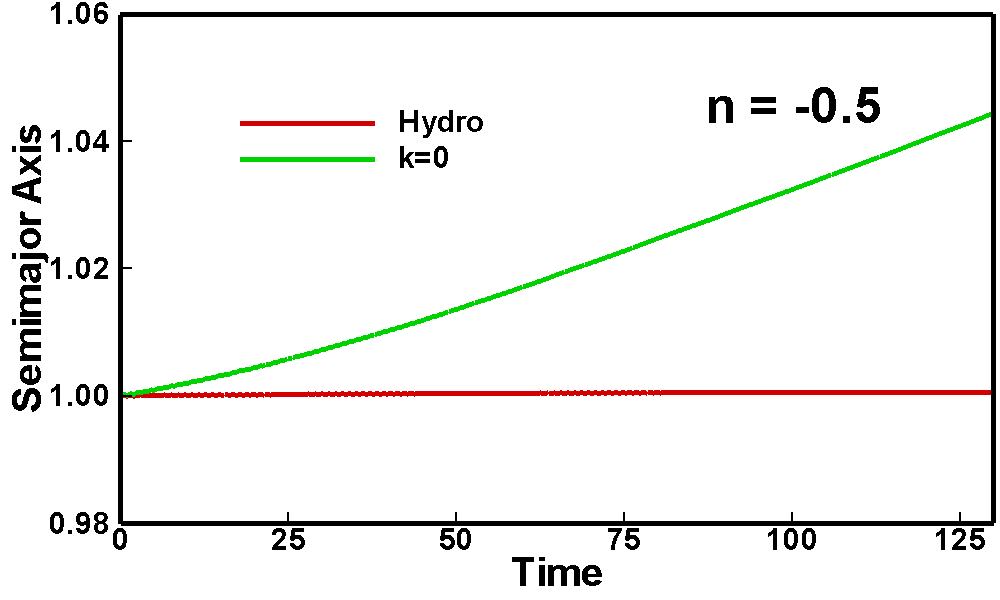}
    \caption{Change in semimajor axis versus time for a $5 M_{\oplus}$ planet embedded in an MHD disc
                  for different values of $k$ (where $\mathfrak{B} \propto r^{-k}$) and $\beta_{i} = 2$. The result for a hydrodynamic disc is shown for reference.
                  \textit{Left Panel:} Result shown for $n=0$, corresponding to models H5n0r1, M5n0k0$\beta$2r1, M5n0k1$\beta$2r1, and 
				       M5n0k2$\beta$2r1.
                  \textit{Middle Panel:} Result shown for $n=1$, corresponding to models H5n1r1,  M5n1k0$\beta$2r1,  M5n1k1$\beta$2r1,  
					M5n1k2$\beta$2r1.
                  \textit{Right Panel:} Result shown for $n=-0.5$, corresponding to models H5n-0.5r1 and M5n-0.5k0$\beta$2r1.}
    \label{figure:sax-time-ndiff}
\end{figure*}

According to the theory presented by \citet{Terquem2003}, these waves exert a torque on the planet that can reverse the planetary migration 
direction if the magnetic field distribution is steep enough. We varied the steepness of the magnetic field, $k$ (see Eq. \ref{eqn:init-power-laws}), 
taking $k=0, 1 $ and $2$, and we obtained different migration rates. Fig. \ref{figure:sax-time-ndiff} (left panel) shows that, for a constant magnetic 
field distribution ($k=0$), the planet migrates inwards but more slowly than in the purely hydrodynamic case. When $k=1$, the planet slowly migrates 
outward, while the planet migrates outward more rapidly when $k=2$. These simulations confirm the result presented by \citep{FromangEtAl2005}: 
in discs with a relatively strong azimuthal magnetic field, the magnetic resonances can slow or reverse a planet's migration.


\begin{figure*}
    \centering
    \includegraphics[width=6cm]{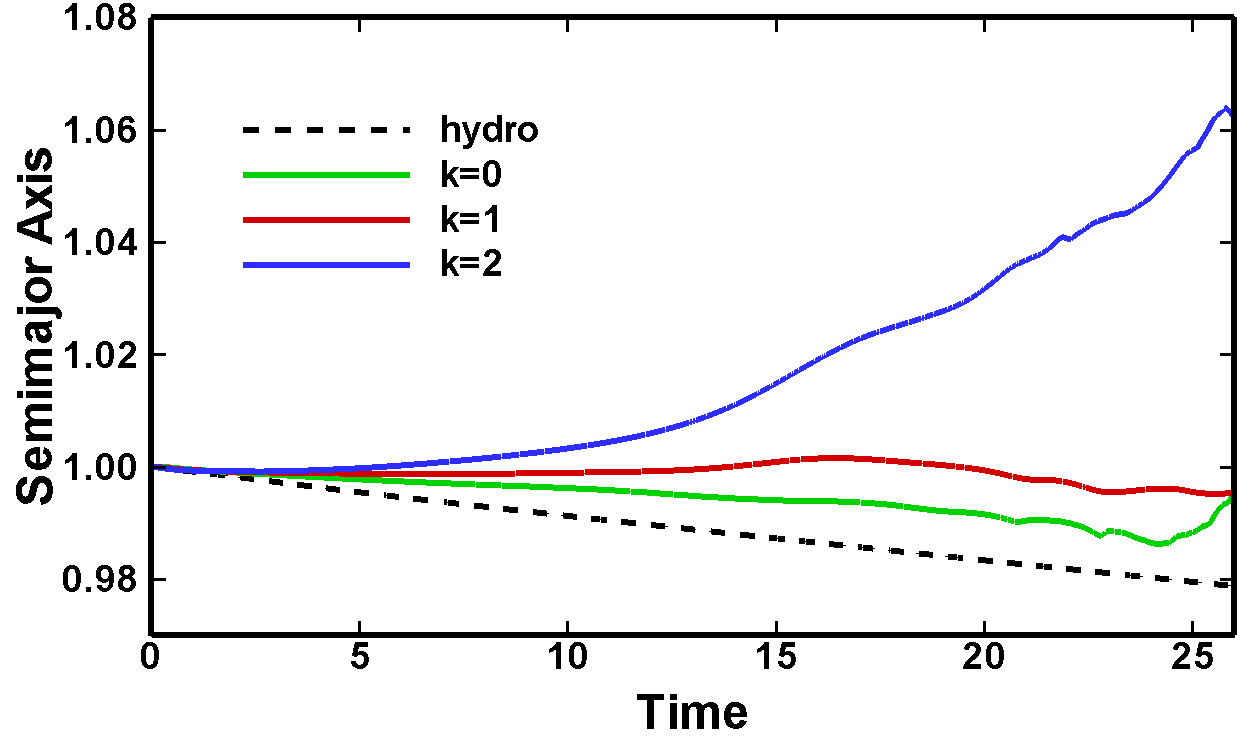}
    \includegraphics[width=6cm]{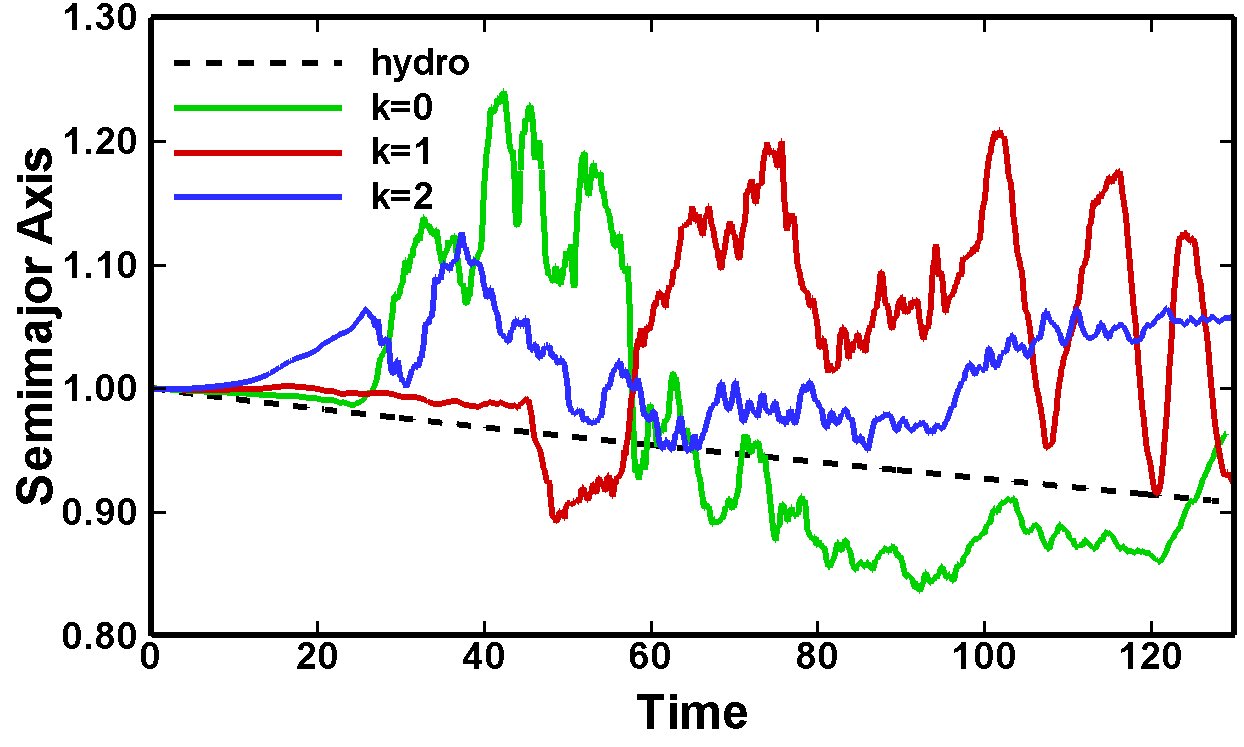}
    \caption{{\it Left Panel:} Variation of the semimajor axis of a $20M_{\oplus}$ planet under the influence of Lindblad and magnetic resonances
				         for various values of initial surface magnetic field exponent, $k$, for the case $n=0$ and $\beta_{i}=2$, 
				         corresponding to models M20n0k0$\beta$2r1, M20n0k1$\beta$2r1, M20n0k2$\beta$2r1, and the hydrodynamic case H5n0r1.
    {\it Right Panel:} Same as in the left panel, but over a longer interval of time.}
    \label{sax-n0k_diff-bet2-20me}
\end{figure*}

\subsection{Migration due to magnetic resonances for different density distributions}

Next, we investigated the action of the magnetic resonances at different surface density distributions. First, we took a density distribution such that 
the density in the disc increases towards the star, $n=1$, and repeated the above simulations at $k=0,1,2$. Fig. \ref{figure:sax-time-ndiff} (middle 
panel) shows that the planet migrates inward in all three cases and, therefore, that the torque from the magnetic resonances is small compared with the 
differential Lindblad torque. The migration rate in the case of $k=1$ almost exactly coincides with the purely hydrodynamic case. For a very steep 
magnetic field distribution, $k=2$, the accretion rate is only slightly slower than for $k=1$.  Overall, we conclude that, in the case of this density 
distribution ($n=1$), the positive torque associated with the magnetic resonances is not strong enough to overcome the negative differential Lindblad 
torque.

In another example, we considered a disc whose density decreases towards the star, $n =-0.5$. This situation is possible, for example, at the inner 
edge of the disc where the expanding magnetosphere or erosion of the disc may push the inner disc away from the star (e.g., \citealt{LovelaceEtAl2008}). When $n=-0.5$, the migration rate in the hydrodynamic case is very low (see Fig. \ref{figure:hydro-5ME-20ME}, top panel) because the negative differential Lindblad torque is approximately compensated by the positive corotation torque. For this surface density distribution ($n=-0.5$), the positive torque associated with the magnetic resonances leads to outward migration of the planet, even for a flat magnetic field distribution, $k=0$ (see Fig. \ref{figure:sax-time-ndiff}, right panel). For a steeper magnetic field distribution, $k=1$, the planet migrates outward even more rapidly. 
It appears, then, that magnetic torques may play a significant role at the disc-cavity boundaries, as well as other regions where the surface density in the disc is flat or decreases towards the star.

We conclude that the rate and direction of migration are determined by the steepnesses of both the magnetic field distribution and the surface density 
distribution. If the surface density \textit{increases} radially towards the star (e.g., $n=1$), then the magnetic resonances do not exert enough torque 
to drive outward migration for any value of $k$. By contrast, when the surface density in the disc \textit{decreases} toward the star (e.g., $n=-0.5$), 
the torque from the magnetic resonances is large enough to drive outward migration for different values of $k$. When the surface density is constant 
(i.e., $n=0$), the magnetic resonances drive outward migration when  $k=1,2$, and inward migration when $k=0$.


\subsection{Migration of a $20 M_{\oplus}$ planet}

The above analysis shows that the effect of an azimuthal magnetic field on a planet's migration strongly depends on both the surface density and magnetic field distributions in 
the disc. It is also interesting to investigate whether the magnetic resonances also depend on the mass of the planet. To investigate this issue, 
we simulated the migration of a $20 M_{\oplus}$ planet in a magnetized disc with a flat surface density distribution ($n=0$) and different steepnesses in the distribution of the magnetic 
field ($k=0, 1, 2$). Fig. \ref{sax-n0k_diff-bet2-20me} (left panel) shows that, for a constant magnetic field in the disc, $k=0$, the migration rate is 
slower than in the hydrodynamic case. For a steeper magnetic field distribution,  $k=1$, the inward migration is even slower, and the direction of the 
migration reverses at an even steeper field distribution, $k=2$. These results are in general agreement with those obtained for a smaller-mass planet: 
the magnetic resonances are strong enough to reverse the migration. Note that, at $k=2$, the outward migration of the more massive planet is much 
faster than the migration of the lower-mass planet.

However, longer simulation runs have shown that, for a $20 M_{\oplus}$ planet, the laminar stage of the disc does not last long, $\lesssim 10-40$ 
planetary orbits. At later times, the disc becomes turbulent. Fig. \ref{sax-n0k_diff-bet2-20me} (right panel) shows that, after a relatively brief stage of 
slow migration in the laminar disc, the migration becomes stochastic in the turbulent disc. We also see such a transition to stochastic migration for a $5 
M_{\oplus}$ planet, though at much later times ($t_{\rm turb} \approx 80$ when $n=0$, $k=0$, and $\beta_{i}=2$). In both cases, the disc becomes 
turbulent, and the semimajor axis varies in time stochastically due to the planet's interaction with turbulent cells in the disc.

We found that the observed turbulence is an interesting phenomenon and investigated the migration of planets in turbulent discs. Earlier, such migration 
was studied by \citet{NelsonPapaloizou2004} in 3D simulations; they observed that the migration becomes stochastic and the migration rate may be 
strongly modified or reversed due to  interaction with turbulent cells in the disc. In this paper, we performed 2D simulations in polar coordinates, and we 
see similar stochastic migration of the planet due to interaction with turbulent cells in the disc. While 2D simulations of the MRI are somewhat restricted, 
they allow us an opportunity to investigate the details of the interaction between the planet and inhomogeneities in the disc. Below, we investigate 
migration in turbulent discs.

%
%

\begin{figure*}
    \centering
    \includegraphics[width=7.5cm]{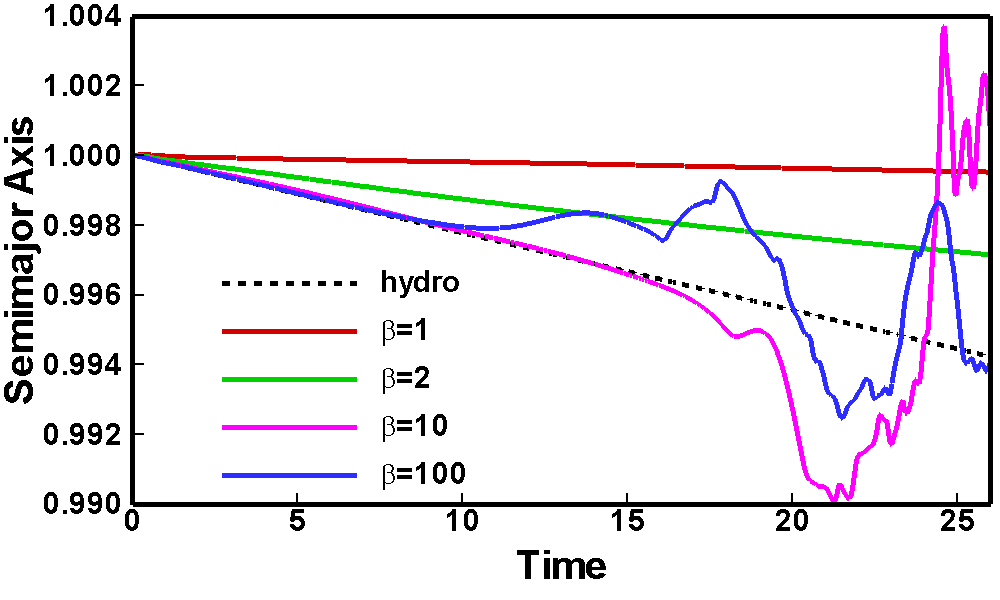}
     \includegraphics[width=7.5cm]{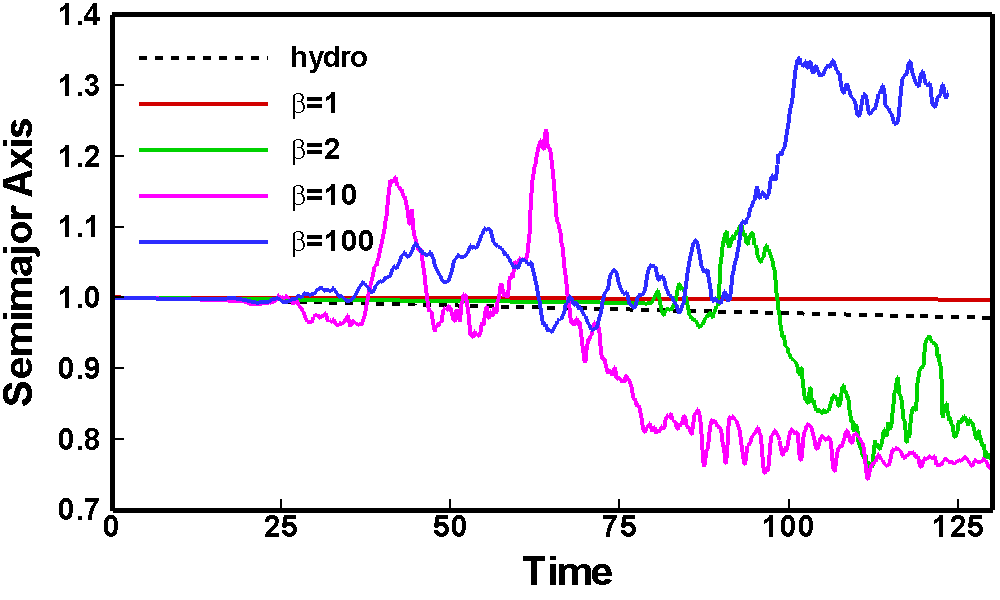}
	\caption{
	{\textit Left Panel:} Variation of the semimajor axis of a $5 M_{\oplus}$ planet for in models with different values of 
				$\beta_{i}$ when $n=0$ and $k=0$, corresponding to models H5n0r1, M5n0k0$\beta$1r1, M5n0k0$\beta$2r1, 
				M5n0k0$\beta$10r1, and M5n0k0$\beta$100r1.
	{\textit Right Panel:} Same as in the left panel, but over a longer interval of time.}
	\label{figure:sax-n0k_diff-bet2-5me}
\end{figure*}

\begin{figure*}
    \centering
    \includegraphics[width=15cm]{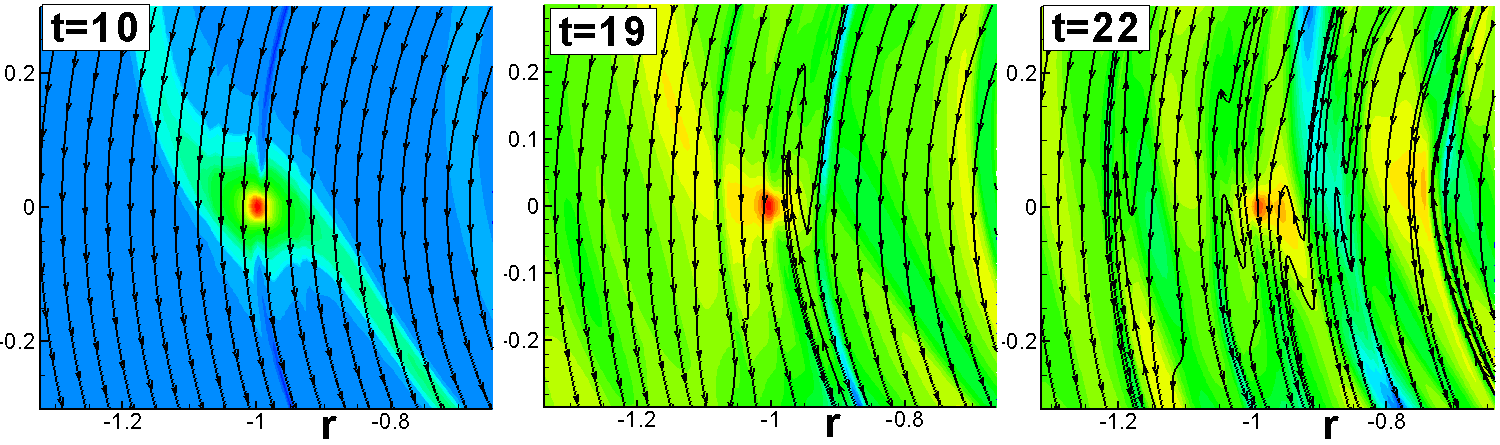}
	\caption{Surface density distribution and magnetic field lines corresponding to model M5n0k0$\beta$100r1 at $t=10, 19, 22$. This figure 
		    demonstrates how initially-azimuthal field lines (left panel) acquire a radial component from non-axisymmetric matter flow near the
                         planet (density waves) and start forming loops (middle panel). Later on, the process spreads to larger distances from the planet, 
                         forming an inhomogeneous distribution of matter and magnetic field in the disc (right panel).}
	\label{figure:mri-startup-3}
\end{figure*}

\begin{figure*}
    \centering
    \includegraphics[width=15cm]{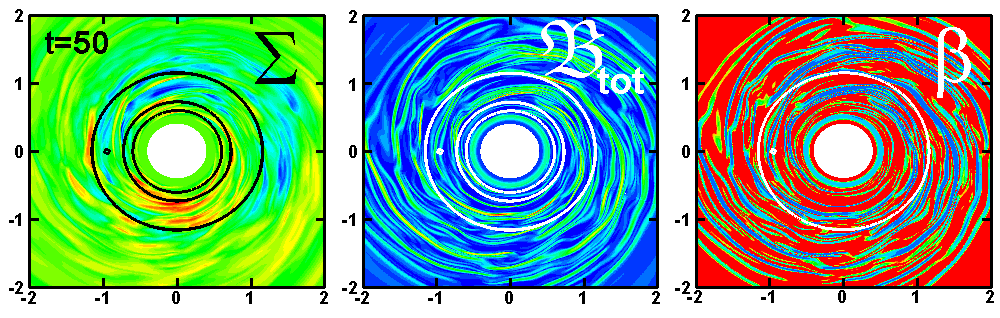}
	\caption{Planet migration in an MRI-turbulent disc in the model M5n0k0$\beta$100r1 at $t=50$.
	\textit{Left Panel:} The surface density variation in the inner part the disc. The planet's location and the $m=1$ and $m=2$ Lindblad resonances 
				are shown via the black circles. 
	\textit{Middle Panel:} The variation of the magnetic field magnitude ($B_{\rm tot}$) in the inner part of the disc. The planet's location and the 
				   $m=1$ and $m=2$ Lindblad resonances are shown via the white circles. 
	\textit{Right Panel:} The variation of the plasma parameter ($\beta_{i}$) in the inner part of the disc. The planet's location and the $m=1$ and 
				 $m=2$ Lindblad resonances are shown via the white circles.}
	\label{figure:mri-analysis-n0k0-bet100-r10}
\end{figure*}

\begin{figure*}
    \centering
    \includegraphics[width=15cm]{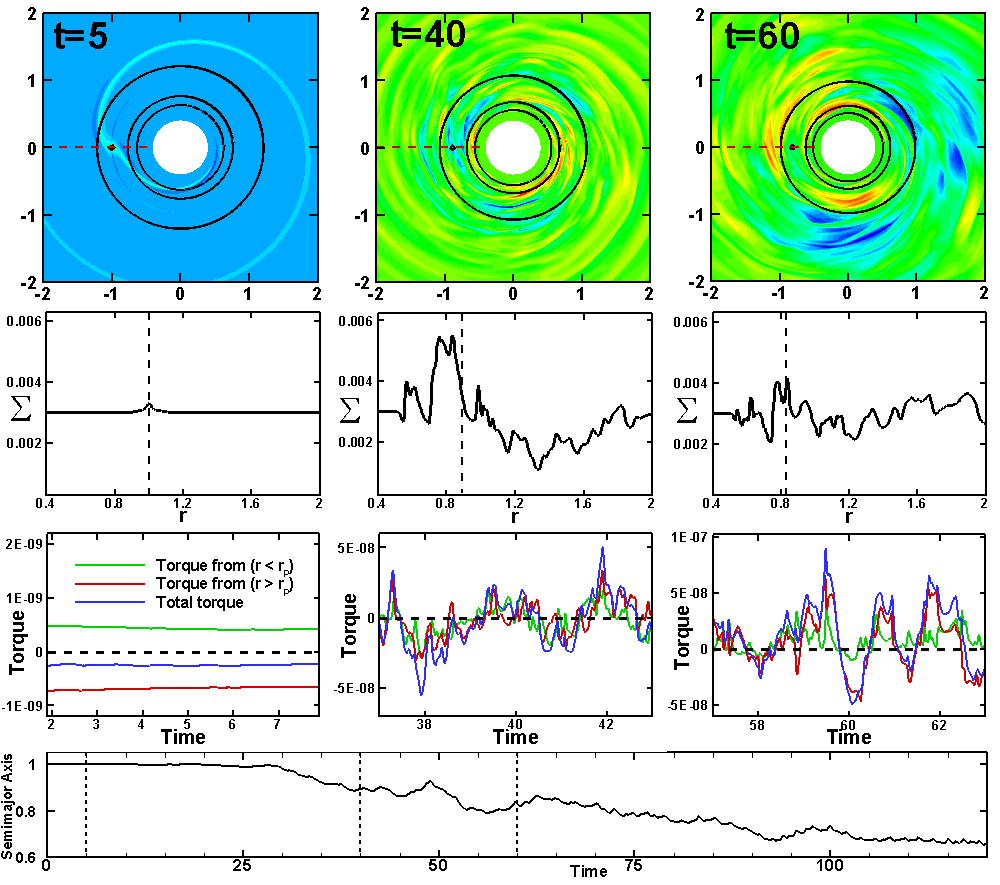}
	\caption{Planet migration in the model M5n0k0$\beta$100r1 shown for $t=5, 40, 60$.
	\textit{Top Panels:} The variation of the surface density ($\Sigma$) in the inner part of the simulation region. The location of the planet, and the 
	 			$m=1$ and $m=2$ Lindblad resonances, are shown via the black circles.
	\textit{Second Row of Panels from the Top:} The one-dimensional variation of the surface density along the red line in the panel above each 
				respective plot. The vertical dashed black line shows the location of the planet.
	\textit{Third Row of Panels from the Top:} The torques acting on the planet from the disc. The green line shows the torque on the planet from 
				the disc where $r < r_{\rm p}$, the red line shows the torque on the planet from the disc where $r > r_{\rm p}$, and the 					blue line shows the total torque on the planet. The dashed black line marks zero torque.
	\textit{Bottom Panel:} The change in the semimajor axis of the planet over time.}
	\label{figure:mri-bet100-torques-sax-10f}
\end{figure*}

\section{Migration in a Turbulent Disc}
\label{sec: migration turbulent}

In this section, we investigate the transition from the laminar to the turbulent disc, the formation of turbulent discs, and the migration of a low-mass 
$5M_{\oplus}$ planet in discs with different strengths of the magnetic field. According to the theory of the MRI (see Sec. \ref{sec:condition for mri}), 
instability is expected in magnetized discs with $\beta_{i}>1$; that is, in discs in which the Alfv\'en velocity $v_A=B/\sqrt{4\pi\rho}$ is smaller than 
the sound speed $c_s$ (e.g., \citealt{BalbusHawley1991,BalbusHawley1998,TerquemPapaloizou1996}). Below, we show the results of simulations at different initial values 
of $\beta_{i}$ (see Sec. \ref{sec:migration turb beta}) and the details of migration in the turbulent disc (see Sec. \ref{sec:migration in turb disc}).

\subsection{Migration in a turbulent disc with different $\beta_{i}$}
\label{sec:migration turb beta}

We studied the migration of the planet in discs with different initial plasma parameters,  $\beta_{i}=1, 2, 10, 100$. 
Fig. \ref{figure:sax-n0k_diff-bet2-5me} (left panel) shows that the disc is initially laminar, and the planet migrates smoothly. This period of smooth 
migration is longest when $\beta_{i}=1$ and $2$, but it is shorter for $\beta_{i}=10$ and even shorter at $\beta_{i}=100$. This is understandable, 
because the magnetic field is more easily tangled by the disc matter when the matter strongly dominates, such as when $\beta_{i}=10$ and $100$. The 
right panel of the same figure shows the semimajor axis of the planet at later times. One can see that the disc is still laminar when $\beta_{i}=1$. In all 
other cases, the disc becomes turbulent. Note that the migration often changes direction from inward to outward and vice versa; this reflects the 
stochastic nature of the migration process in a turbulent disc, as discussed below in Sec. \ref{sec:migration in turb disc}.


\subsection{Migration in a turbulent disc with $\beta_{i}=100$}
\label{sec:migration in turb disc}

In this section, we consider the development of the MRI, as well as the migration of a planet in a turbulent disc, in greater detail. As a base, we 
consider a model where the initial value of plasma parameter is large, $\beta_{i}=100$ (model M5n0k0$\beta$100), so that the MRI instability starts 
easily.

We observed that the origin of the radial component of the field, which is required for the instability,  is in the fact that the planet excites 
non-axisymmetric waves in the disc. These waves lead to non-axisymmetric motion of the gas in the disc and, subsequently, to the formation of a radial 
component of the magnetic field, which is further stretched by the differential rotation in the disc. Fig. \ref{figure:mri-startup-3} shows how parts of the initially azimuthal magnetic field near the planet (left panel) 
acquire a radial component and are subsequently stretched by the differential rotation in the disc, eventually forming a loop (middle panel). Later, this 
process occurs at larger distances from the planet, many more field lines are stretched in the radial direction (such that different inhomogeneities 
and loops form), and the disc becomes globally inhomogeneous and turbulent.

Fig. \ref{figure:mri-analysis-n0k0-bet100-r10} (left panel) shows that the disc consists of azimuthally-stretched turbulent cells. The middle panel of the 
same figure shows that the total magnetic field, $B_{\rm tot}$, becomes strongly inhomogeneous, with some regions having the original polarity and others reversing
polarity. The distribution of the plasma parameter, $\beta_{i}$, shows that the simulation region splits into regions that are either magnetically- or 
matter-dominated (right panel). Therefore, in the MRI regime, the disc becomes strongly inhomogeneous both in the density and in the magnetic field.

We investigated the density distribution, torque, and semimajor axis evolution of the planet in this case in greater detail.  
Fig. \ref{figure:mri-bet100-torques-sax-10f} (top panels) shows the surface density distribution in the disc at $t = 5, 40$ and $60$. At $t=5$, the disc is still laminar and two Lindblad density waves are clearly seen. However, after $t\approx10$, non-axisymmetric motions start to ``tangle'' the field lines of the weak magnetic field, and an MRI-type turbulence gradually develops. MRI-type turbulence is observed during the time interval $10 \lesssim t \lesssim 60$. 

The middle column ($t=40$) shows small-scale turbulent cells that persist for many orbits. However, later 
(at $t = 60$), the turbulent cells become larger in size because ``islands'' of stronger magnetic field also become larger in size, and often one or two 
main density waves form in the inner parts of the disc. The beginning of this process is seen in the right column of 
Fig. \ref{figure:mri-bet100-torques-sax-10f}. We suggest that the finite life of the MRI turbulence and formation of these larger-scale waves may be 
connected with the 2D nature of our MRI turbulence. However, the low-amplitude MRI turbulence proceeds over long periods of time, which is sufficient 
to study the migration of a planet in the turbulent disc. Formation of large-scale waves is possible in realistic discs; we use these waves to 
study the interaction of a planet with waves in the disc in Sec. \ref{sec:migration in waves}.

The second row in Fig. \ref{figure:mri-bet100-torques-sax-10f} shows the 1D density distribution along the line connecting the planet and the center of the star at 
the same times as the top row. When the disc is laminar, we see only a small bump in density associated with matter accumulation near the planet. 
However, as the disc becomes more and more turbulent, larger and larger variations in the surface density of the disc are observed. 

The 3rd row in Fig. \ref{figure:mri-bet100-torques-sax-10f} shows the torques acting on the planet at the same times as the top two rows. When the 
disc is laminar (left column), the planet migrates due to the excitation of density waves at the Lindblad resonances: the inner torque is positive and 
smaller than the negative outer torque, so that the total torque is negative (the blue line in Fig. \ref{figure:mri-bet100-torques-sax-10f}). In this case, 
the planet slowly migrates inward (see the slow variation of the semimajor axis up to time $t\approx 30$ in the bottom row). However, when the disc is 
turbulent ($t=40, 60$), we observe that both the inner and outer torques can be either positive or negative; they both 
vary rapidly  and the resulting torque also changes sign.

The \textit{magnitudes} of these turbulent torques are much larger than those seen in the laminar case (note the torque magnitudes shown on 
the $y$-axes in the third row of Fig. \ref{figure:mri-bet100-torques-sax-10f}). The torques become stochastic and correspond to the interaction of the 
planet with individual turbulent cells. The variation of the sign of the net torque shows also that the total averaged torque may be either negative or 
positive (i.e., that the direction of migration may change). In the shown example, the average total torque is negative and the planet migrates inward 
overall.  Note that, in other cases, the direction of  the migration may be either inward or outward (see, e.g., Figs. \ref{sax-n0k_diff-bet2-20me} and 
\ref{figure:sax-n0k_diff-bet2-5me}).

%
%

\section{Interaction between a planet and waves in the disc}
\label{sec:migration in waves}

\begin{figure*}
    \centering
    \includegraphics[width=15cm]{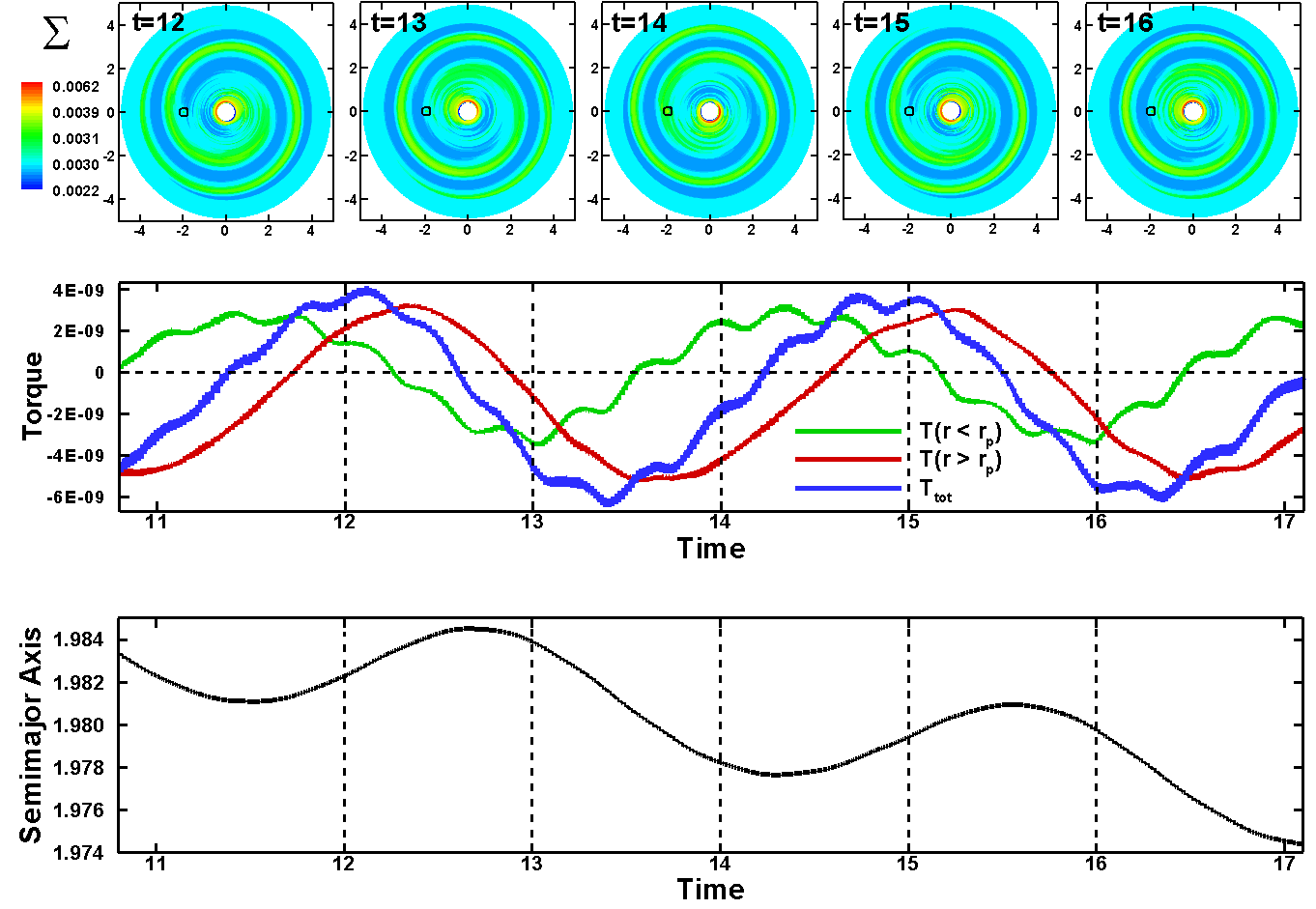}
	\caption{Migration of a $5 M_{\oplus}$ planet in a hydrodynamic disc in the model W5n0r2, for which low-amplitude ordered density waves are 
		    propagated through the disc, at $t = 12, 13, 14, 15, 16$. The bottom two panels mark these times with vertical black dashed lines.
	\textit{Top Panels:} The surface density ($\Sigma$) variation in the disc; the small black circle shows the location of the planet.
	\textit{Middle Panel:} The torque on the planet from the disc.  The green line shows the torque on the planet from the disc where 
				   $r < r_{\rm p}$; the red line shows the torque on the planet from the regions of the disc where $r > r_{\rm p}$, and the blue line shows the 
				   total torque on the planet. The horizontal dashed black line shows zero torque.
	\textit{Bottom Panel:} The change in the semimajor axis of the planet over time.}
	\label{figure:wave-weak-rp2}
\end{figure*}

\begin{figure*}
    \centering
    \includegraphics[width=15cm]{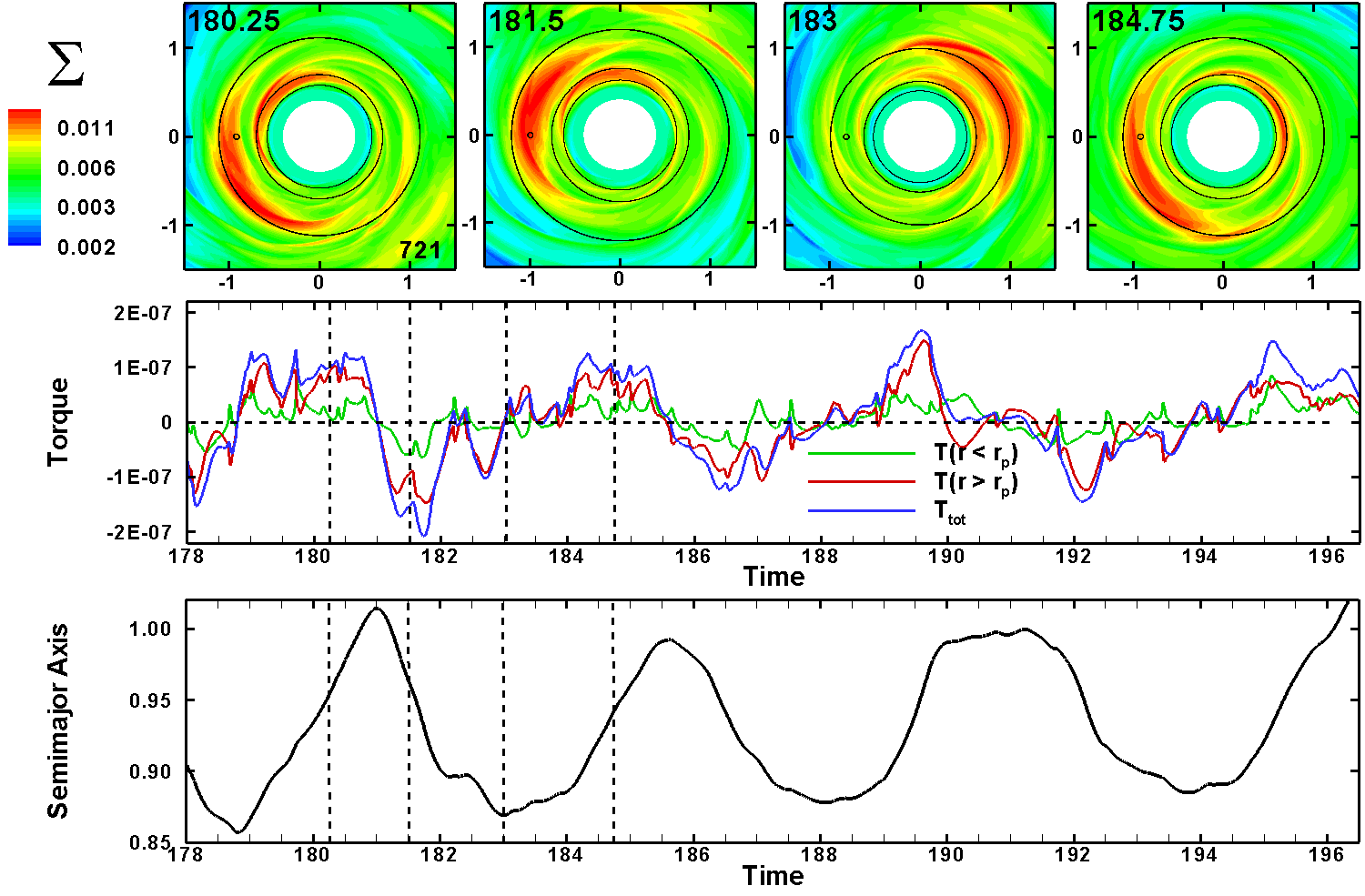}
    \caption{Planet migration in high-amplitude waves in the model W5n0k0$\beta$100r2 for $t=180.25, 181.5, 183,184.75$. In the two bottom
                  panels, these times are shown by vertical black dashed lines.
                  \textit{Top Panels:} Surface density variation in the inner part of the simulation region. The large black circles show position of the Lindblad 
			                   resonances, while the small black circle shows the position of the planet.
                  \textit{Middle Panel:} The torque on the planet from the disc. The green line shows the torque on the planet from the part of the disc where 
					$r < r_{\rm p}$; the red line shows the torque on the planet from the disc where $r > r_{\rm p}$. The blue line shows 
					the total torque on the planet.  The horizontal dashed black line shows zero torque.
                  \textit{Bottom Panel:} The variation of the semimajor axis of the planet over time.}
    \label{waves-rp2-bet100-5me}
\end{figure*}

The interaction between a planet and turbulent cells in a disc is a complex process. The planet interacts with a set of turbulent cells gravitationally, but the 
Lindblad density waves are not homogeneous. The closest cell may contribute to the torque more strongly than more remote cells, and the torque 
becomes more stochastic. Additionally, the planet passes through individual cells, and each cell may exert a corotation torque on the planet. It 
is difficult to track the interaction of a planet with individual turbulent cells. That is why we developed conditions in which a planet interacts with ordered 
inhomogeneities (in the form of waves in the disc). We consider two types of waves: (1) low-amplitude ordered waves generated in a hydrodynamic disc 
by a force at the inner boundary (Sec. \ref{subsec:low-amp waves}), and (2) high-amplitude waves that form at later times in simulations using an MHD 
disc (Sec. \ref{subsec:high-amp waves}.)

\subsection{Interaction between a planet and low-amplitude waves in a hydrodynamic disc}
\label{subsec:low-amp waves}

To better understand the interaction between a planet and individual turbulent cells, we created a model in which an ordered density wave is generated 
at the inner disc boundary by a periodic force that decreases with the radius as $r^{-3}$.  This force generates density waves with a small amplitude, for which the density 
contrast between the wave and the disc is  small, about $5-7\%$. The density wave then propagates through the simulation region. We placed a 
planet at an initial radius of $r_{{\rm p},i}=2$, away from the action of this force at the boundary.

We observed that the planet moves faster than the wave and it interacts differently with different parts of the wave. Fig. \ref{figure:wave-weak-rp2} 
(top panels) shows slices of the surface density at $t=12,13,14,15,16$. We show the moments when the planet is located at the inner or outer edge of 
a wave (i.e., where the surface density either decreases or increases toward the star, respectively). The corotation torque is larger than the differential 
Lindblad torque when the density slope in the disc corresponds to that in Equation (\ref{eqn:gamma-total}); that is, the slope of the density distribution 
is either positive or only slightly negative toward the star (see Sec. \ref{sec:migration hydro}). This is expected when the planet is at the inner edge of 
a wave.

The middle row of Fig. \ref{figure:wave-weak-rp2} shows the torques acting on the planet; the times from the top panels marked with vertical dashed 
lines. The total torque has maxima at $t=12$ and $t~=~15$. The top panels show that, at these moments of time, the planet is located at the inner 
edge of the wave and, therefore, that the positive torque acting on the planet is the corotation torque (which appears due to the positive density 
gradient at the inner edge of the wave). The bottom panel of Fig. \ref{figure:wave-weak-rp2} shows that the planet migrates inward overall, because 
the differential Lindblad torque dominates overall. However, at $t=12$ and $t=15$, the planet migrates outward due to the temporarily dominant 
corotation torque.

At $t=13$ and $t=16$, the planet is located at the outer edge of the density wave, where the density increases towards the star. At these moments of 
time, the total torque is negative and the planet migrates inward. We suggest that, at the outer edge of a wave, a planet excites density waves at the 
Lindblad resonances, and the differential Lindblad torque drives overall inward migration. However, when the planet is at the inner edge of the wave, 
the positive corotation torque is large, and the planet migrates outward.

\subsection{Interaction between a planet and high-amplitude waves in an MHD disc}
\label{subsec:high-amp waves}

In this subsection, we analyze the passage of a planet through waves with much higher amplitudes. These high-amplitude  waves often form at later 
times in simulations of MRI-turbulent discs, where the non-axisymmetry of the gravitational potential leads to the formation of the inner density waves. We 
chose the model where the initial plasma parameter $\beta_{i}~=~100$ and a planet is placed at $r_{{\rm p},i}=2$. At later moments in time, after the 
wave forms, the planet migrates to a radius of $r_{\rm p}\approx 0.9-1$. Fig. \ref{waves-rp2-bet100-5me} (top panels) shows the density distribution 
in the wave and the position of the planet for several representative moments in time, where the planet is located at the inner edge of a wave (left and 
right panels, $t=180.25, 184.75$), the planet is in the middle of a wave ($t=181.5$), and the planet is in the low-density region ($t=183$). The 
density contrast between the wave and the rest of the disc is $\sim 70\%$, which is about $10$ times higher than the low-amplitude waves discussed in 
the previous subsection.

The middle and bottom rows of Fig. \ref{waves-rp2-bet100-5me} show the torques acting on the planet and the variation of the planet's semimajor axis 
in time. The dashed vertical lines show the four moments in time corresponding to the top four panels. One can see that, at $t=180.25$, the planet is 
located at the inner edge of the wave, where the density decreases towards the star, and the positive corotation torque is expected to be larger than 
the differential Lindblad torque. Indeed, the middle row shows that the total torque is positive at this moment of time, and the planet migrates outward. 

At the second moment in time, $t=181.5$, the planet is located in the middle of the density wave. The middle row shows that, at this moment, a 
strongly negative torque dominates, and the planet migrates inward. This large negative torque is primarily comprised of the differential Lindblad torque. This torque is proportional to the surface density in the disc, and therefore it is large when the planet is located 
inside the high-density wave.  The positive corotation torque is relatively small. At $t=183$, the planet is located away from the wave in the low-density part of the disc, and both torques are small. At the last considered moment ($t=184.75$), the planet is again at the inner edge of the wave, where the corotation torque is positive and larger than the differential Lindblad torque, and the planet migrates outward.

In this example, the density wave is an analog of a large turbulent cell, where the total torque is either positive or negative and acts onto the planet 
depending on the position of the planet relative to the wave. We expect that the interaction with smaller-sized turbulent cells is similar to the observed 
interaction with MHD waves, but that the duration of the interaction is shorter. As a result of such interactions, a torque is exerted on the planet during 
short intervals of time, and the planet's semimajor axis varies stochastically under the action of these torques.

\begin{figure}
    \centering
    \includegraphics[width=\columnwidth]{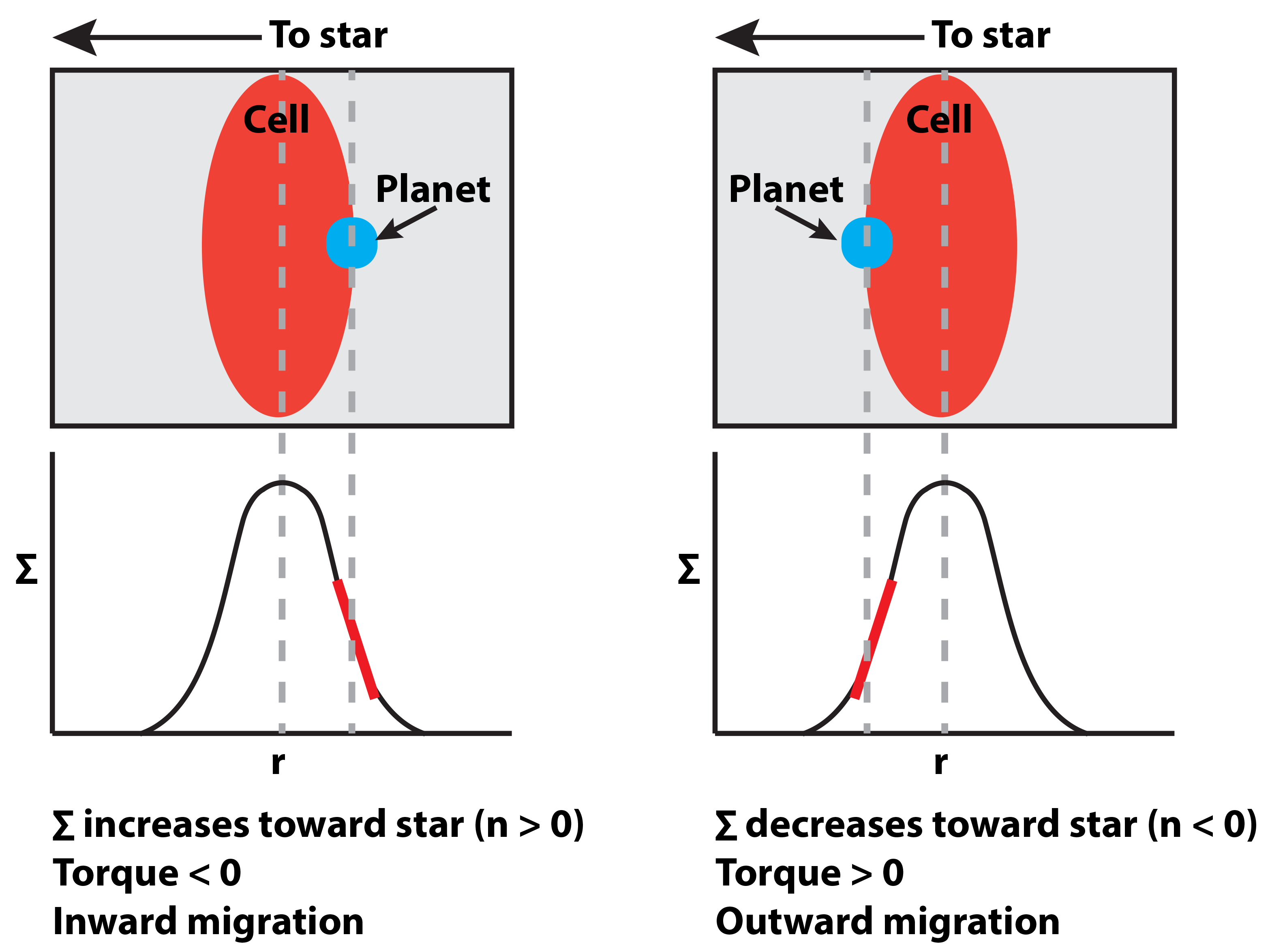}
    \caption{Schematic showing the interaction between a planet and a turbulent cell.}
    \label{fig:planet-cell}
\end{figure}

Based on our observations with both the low- and high-amplitude waves we can schematically describe the interaction between a planet and a turbulent cell (see \fig{fig:planet-cell}). When the planet is at the outer edge of the cell (the side that is farther from the star), the density is increasing toward the planet ($n > 0$), and so we expect the torque to be negative and the planet to migrate inward. Conversely, when the planet is on the inner edge of the cell (the side that is closer to the star), the density decreases toward the star ($n < 0$), and so we expect the torque to be positive and the planet to migrate outward. This inward-outward migration does not ``trap'' the planet, because the cell is moving away from the star, so this is a transient process.

%
%

\section{Conclusions}
\label{sec:conclusions}

We investigated the migration of a low-mass planet ($5M_\oplus$) in magnetized discs using a Godunov-type HLLD MHD code in polar coordinates \citep{KoldobaEtAl2015}. The initial surface magnetic field is azimuthal, with different radial distributions $B_{\varphi}\sim r^{-k}$, and different strengths determined by the initial plasma parameter at the planet's location, $\beta_{i}$, which varied between $\beta_{i}=1$ and $100$. We also varied the initial radial surface density distribution in the disc, $\Sigma\sim r^{-n}$. Our main conclusions are as follows:

\begin{enumerate}

\item In strongly-magnetized discs ($\beta_{i}=1, 2$), and where the density distribution in the disc is flat ($n=0$), the planet's migration is strongly 
influenced by magnetic resonances, which are excited near the planet and exert a positive torque on the planet. The migration slows down when the 
magnetic field distribution is flat ($k=0$); it slows down more strongly when the magnetic field increases towards the star ($k=1$); 
and the migration  reverses when the field steeply increases towards the star ($k=2$). These results are in accord with theoretical predictions by 
\citet{Terquem2003} and simulations by \citet{FromangEtAl2005}.

\item Compared with \citet{FromangEtAl2005}, we investigated the effect of magnetic resonances on the migration of a planet in discs with different 
density distributions. We observed that the steepness of the density distribution strongly influences the rate and direction of the migration. When the 
density increases towards the star ($n=1$) the planet migrates inward at any steepness of the magnetic field ($k=0, 1, 2$) and the effect of the 
magnetic resonances is negligibly small. This is because, at larger density steepness, the negative differential Lindblad torque is much larger than the 
positive corotation or magnetic torques. In the opposite situation, when the density in the disc decreases toward the star ($n=-0.5$), and the positive 
corotation torque almost balances the negative Lindblad torque, the role of the positive magnetic torque becomes very significant. The planet migrates 
outward due to magnetic resonances at all values of the magnetic field steepness ($k=0,1,2$).

\item Experiments with a larger mass planet, $20 M_{\oplus}$, show that the action of the magnetic resonances is similar to that in the case of a lower-mass planet: the positive torque increases with the steepness of the field, as predicted by the theory. We also observed that, at $k=2$, the more 
massive planet migrates outward more rapidly than the lower-mass planet. This is in accord with Eqn. (65) of \citet{Terquem2003}, in which the 
migration time scale is shown to be $\propto M_{\rm p}^{-1}$.

\item We investigated weakly-magnetized discs with initial plasma parameter $\beta_{i}=10-100$. We observed that non-axisymmetric motions in 
the disc lead to the formation of a radial component of the magnetic field, its stretching by the differential rotation in the disc, and subsequent MRI-
driven turbulence in the disc. Interaction between the planet and turbulent cells leads to stochastic migration, similar to that observed in 3D discs by, 
e.g., \citet{NelsonPapaloizou2004}. We investigated the transition from the laminar to turbulent disc that often starts near vicinity of the planet, where 
non-axisymmetric density waves are excited by the planet. The turbulence starts more rapidly when the planet is more massive.

\item The torques acting on the planet are larger in turbulent discs than in the laminar discs, leading to more rapid migration. However, the direction of 
the migration is also stochastic, and outward migration is frequently observed.

\item To understand how planets interact with individual turbulent cells, we investigated the propagation of a planet through density waves in the disc. 
We observed that a planet experiences a strong negative torque when it is located inside the wave or at the outer edge of the wave (where the  
density increases towards the star). However, when the planet is located at the inner edge of the wave (where the density decreases towards the 
star), then it experiences a positive corotation torque, and it migrates away from the star. We conclude that the stochastic motion of the planet is 
connected with the alternating action of these positive and negative torques.

\end{enumerate}

%
%

\section*{Acknowledgments} 
We gratefully acknowledge support from the NASA Research Opportunities in Space and Earth Sciences (ROSES) Origins of Solar Systems grant NNX12AI85G. We also acknowledge support from grants FAP-14.B37.21.0915, SS-1434.2012.2, and RFBR 12-01-00606-a. AVK was supported by the Russian academic excellence project ``5top100". Resources supporting this work were provided by the NASA High-End Computing (HEC) Program through the NASA Advanced Supercomputing (NAS) Division at the NASA Ames Research Center and the NASA Center for Computational Sciences (NCCS) at Goddard Space Flight Center.

%
%

\bibliographystyle{mn2e}

\end{document}